# A machine-compiled macroevolutionary history of Phanerozoic life

**Shanan E. Peters[1], Ce Zhang[2], Miron Livny[2], Christopher Ré[3]**

[1]Department of Geoscience, University of Wisconsin-Madison, Madison, WI, 53706 USA. [2]Department of Computer Science, University of Wisconsin-Madison, Madison, WI, 53706 USA. [3]Department of Computer Science, Stanford University, Stanford, CA 94305 USA.

Corresponding author: Shanan E. Peters, Department of Geoscience, University of Wisconsin-Madison, 1215 W. Dayton St., Madison, WI, 53706; phone: 608-262-5987, email: peters@geology.wisc.edu

**Many aspects of macroevolutionary theory and our understanding of biotic responses to global environmental change derive from literature-based compilations of palaeontological data. Existing manually assembled databases are, however, incomplete and difficult to assess and enhance. Here, we develop and validate the quality of a machine reading system, PaleoDeepDive, that automatically locates and extracts data from heterogeneous text, tables, and figures in publications. PaleoDeepDive performs comparably to humans in complex data extraction and inference tasks and generates congruent synthetic macroevolutionary results. Unlike traditional databases, PaleoDeepDive produces a probabilistic database that systematically improves as information is added. We also show that the system can readily accommodate sophisticated data types, such as morphological data in biological illustrations and associated textual descriptions. Our machine reading approach to scientific data integration and synthesis brings within reach many questions that are currently underdetermined and does so in ways that may stimulate entirely new modes of inquiry.**

Palaeontology is based on the description and classification of fossils, an enterprise that has played out in an untold number of publications over the past four centuries. The construction of synthetic databases that aggregate fossil data in a way that enables large-scale questions to be addressed has expanded the intellectual reach of palaeontology (1-5) and led to fundamental new insights into macroevolutionary processes (e.g., 6-9) and the timing and nature of biotic responses to global environmental change (e.g., 10,11). Nevertheless, palaeontologists often remain data limited, both in terms of the pace of discovery and description of new fossils and in terms of their ability to synthesize existing knowledge on the fossil record. Many other sciences, particularly those for which physical samples and specimens are the source of data, face similar challenges.

One of the most successful efforts to compile data on the fossil record to date is the Paleobiology Database (PBDB). Founded nearly two decades ago by a small team who generated the first sampling-standardized global Phanerozoic biodiversity curves (12,13), the PBDB has since grown to include an international group of more than 380 scientists with diverse research agendas. Collectively, they have spent approximately nine continuous person years entering almost 300,000 taxonomic names, 518,000 opinions on the status and classification of those names, and 1.2 million fossil occurrences (i.e., temporally and geographically resolved instances of fossils). Some data derive from the original fieldwork and taxonomic studies of the contributors, but the majority of the data were extracted from some 40,000 publications. Nevertheless, the PBDB leverages a small fraction of all published palaeontological knowledge, primarily because there is a large and ever-growing body of published work and manually entering data is a labor intensive and often ambiguous task. Moreover, because the end product of manual data entry is a list of facts that are divorced from most, if not all, original contexts, assessing the quality of the database and the reproducibility of results is difficult.

Here we develop and deploy PaleoDeepDive (PDD), a statistical machine reading and learning system, to automatically find and extract fossil occurrence data from the scientific literature. Our motivations for doing so are threefold. First, we aim to test the reproducibility of several key macroevolutionary results that are used to frame much of our understanding of the large-scale history of life (1-13). Second, we aim to improve upon the state of the art in machine reading systems, which have not been deployed and validated in a result-focused scientific application. Third, we aim to develop a system that has the capacity to change the practice of science by removing substantial time and cost barriers to large-scale data integration and synthesis. In so doing, we hope to shift the balance of effort away from slow and expensive data compilation efforts and towards creative hypothesis testing and the more focused and efficient generation of new primary data.

The specific question that motivates this study is: Can the data produced by a machine reading system achieve a quality that is sufficient to enable literature synthesis-based science? We address this question by pitting our system's results against those of human-constructed databases at several levels of granularity, from individual facts that describe opinions on the biological classification of taxa to synthetic results that summarize the history of genus-level biodiversity over millions of years. In all cases, we show that PDD produces data with quality that is at least as good as that generated by humans, even when only small amounts of training data are available. We also test the ability of our system to lower the cost of extending a human-constructed database by extracting data from an order of magnitude more references. The results of this experiment show that our system is efficiently scalable and that key macroevolutionary patterns are robust even when derived from different bodies of literature. We further test the ability of our system to incorporate new types of information by extracting morphological data from biological illustrations and their labels, captions, and associated text. Our machine-derived body size estimates are statistically indistinguishable from those produced by humans manually measuring the same illustrations. Because our system is designed for broad applicability in the biological and physical sciences, it can be readily extended for knowledge base creation in many different domains of Earth and life science.

**System Description**

**Overview.** A fundamental challenge faced by machine reading systems is that computers cannot read documents unambiguously. Instead, machines have difficulty with all aspects of document reading, from optical character recognition (OCR) and natural language understanding tasks, to the more complex subtleties involving domain-specific representations of fact. As a result, coping with ambiguity is a key challenge in many areas of computer science (14–18).

To accommodate the inherent ambiguity of the scientific literature, PDD is built upon the DeepDive machine reading infrastructure (18), which is designed to extract information in a way that achieves a deep level of contextual understanding. To do this, DeepDive takes a radical approach: it treats all sources of information, including existing data, as evidence that may or may not be correct. Extraction tasks then become probabilistic inference challenges. DeepDive takes a joint or collective probabilistic approach (19), in which all available information is considered simultaneously. This is in contrast to a pipelined approach (17, 20, 21), in which hard decisions are made after each stage of document processing, which can result in compounding errors and suboptimal data quality (22). DeepDive is also able to accept diverse forms of feedback, including example data sources, formal rules, and training data.

Similar conceptual underpinnings are currently in use by Google's Knowledge Graph, IBM's Watson, and CMU's NELL project. However, none of these have demonstrated an ability to extract information collectively from text, tables, and figures, which is critical to meeting the standards and questions posed by scientific uses. The cost of a collective probabilistic approach is that complexity grows exponentially with each new source of ambiguity. Recent work, in part motivated by this study, allows us to perform the requisite statistical inference tasks orders of magnitude more efficiently than was possible just a few years ago (23-27).

**PaleoDeepDive Pipeline.** The input to PaleoDeepDive is a set of documents (e.g., PDFs or HTML), and a database structure that defines entities and relationships of interest. The first step in the DeepDive process is to perform document parsing, including optical character recognition (OCR), document layout recognition, and natural language processing (NLP) of the text (Supplementary Fig. 1). These steps are required before applying any of the reasoning necessary to recognize entities and the relationships among them. An example of the latter is: "Does this instance of the word 'Waldron' refer to the 'Waldron Shale', a distinct geological formation, and if so, what is its stated geologic age, where is it located geographically, and which species are reported from it?" Descriptions of how to recognize entities and the relationships among them can be articulated by scientists through rules and examples (Supplementary Fig. 2; Supplementary Tables 1, 2). The weights of these rules are then estimated (i.e., learned) from the data using classical equations based on exponential models (19). Essentially, the likelihood of the given set of observations (data and rules) is maximized, given the set of features expressed by the rules (Supplementary Fig. 3).

The end-product of PDD is not a classical database, which consists of isolated facts that are all assumed to be equally correct. Instead, DeepDive produces a probabilistic database in which each fact remains tightly coupled to its original context and is associated with an estimated probability of being correct (28). A probabilistic approach is not a panacea, but it does allow our system to cope with ambiguity in a principled and consistent way. This is critical for scientists, who can use these probabilities to identify errors and omissions and thereby improve the quality of the system. For further explanation, application code, and example data output see our online documentation (http://deepdive.stanford.edu/doc/paleo.html).

**Results**

**Overlapping Document Set (ODS).** To quantitatively assess PDD's ability to read the literature and extract structured fossil occurrence data, we used the human-constructed PBDB as a baseline for comparison. Specifically, 11,782 documents from the top-50 serial publications in the PBDB were also accessible to us and processed by PDD (Supplementary Table 3). This experiment allows comparisons to be made between human readers and our system at every level of granularity. Because PDD depends on linguistic understanding, our system is currently able to process English, German, and Chinese language documents, which constitute 76%, 6%, and 2%, respectively, of PBDB's total reference inventory.

On average, PDD extracts more taxonomic data from a document than humans. For example, humans extracted 79,913 opinions on the status and biological classification of taxonomic names from the ODS, whereas PDD extracts 192,365 opinions. Although many of these opinions are simple cases that are often not entered by humans (e.g., a species belongs to a genus), they nonetheless constitute taxonomic information which is sometimes not entered by humans at all. For example, PDD extracted 59,996 taxonomic names from the ODS that were never formally entered by human readers from any of the over 40,000 references they have entered thus far. A random sample of these names indicates that most are valid species-level taxa and that ≥90% were correctly extracted (Supplementary Table 4). The cases where PDD fails to recognize and extract data from a document are due primarily to OCR-related errors (Supplementary Tables 5, 6), which are orthogonal to this work.

The quality of PDD's database was assessed in three ways. The first uses DeepDive's pipeline, which produces internal measures of precision for every entity and relationship. All of the extractions used here have a precision of ≥ 95% according to this criterion. We also conducted blind assessment experiments of two types. In the first double blind experiment, we randomly sampled 100 relations from the PBDB and PDD and then randomized the combined 200 extractions into a single list. This list was then manually assessed for accuracy relative to the source document. In this assessment, PDD achieves ≥ 92% accuracy in all cases, which is greater than or equal to the accuracy estimated for the human database (Supplementary Table 7). In the second blind experiment, eight scientists with different levels of investment in the PBDB were presented with the same five documents and the same 481 randomly selected taxonomic facts, which were extracted by both humans and PDD (Supplementary Fig. 4). No indication was given regarding which system generated the facts. Humans measured a mean error frequency in the machine-constructed database of 10%, with a standard deviation of ±6%. This is comparable to the error rate of 14 ±5% they estimated for those same documents in the human-constructed database (Supplementary Fig. 5). Variability in estimates between annotators reflects a combination of assessment error and divergent interpretations of the data. Although these blind experiments suggest that the error rate is comparable between the databases, the comparisons are not strictly equivalent. For example, PDD currently understands only parent-child relationships and synonymy, which comprise a large fraction (90% and 5%, respectively) but not all of the taxonomic opinions in the PBDB. Human data enterers also selectively enter data that are deemed important or non-redundant with data in other documents because the data entry process is time consuming.

The third approach we took to assessing PDD quality was conducted at the aggregate level of Phanerozoic macroevolutionary patterns (29). After processing both databases with the same algorithms to generate a working taxonomy and a list of occurrences meeting the same threshold of temporal resolution (i.e., epoch or finer), we find good overall agreement in macroevolutionary results (Fig. 1; data are binned into the same 52 time intervals, mean duration 10.4 Myr). Both long-term trends and interval-to-interval changes in genus-level diversity and turnover rates are strongly positively correlated, indicating that both databases capture the same signal. The number of genus-level occurrences in each time interval, which is important to sampling standardization approaches (30,31), are also positively correlated (for first differences, Spearman rho = 0.65; $p = 5.7 \times 10^{-7}$). The times of first and last occurrence of 6,708 taxonomically and temporally resolved genera common to both database are also congruent (Fig. 2).

Differences between results (Fig. 1) can be attributed to a combination of errors and inconsistencies in the human-constructed database, as well as to data recovery and inference errors committed by PDD. For example, the PBDB contains typographical errors introduced during data entry. But, most of the differences observed in Fig. 1 are attributable to more insidious inconsistencies. For example, there are groups of occurrences in the PBDB that derive from multiple documents, even though only one document is cited. Occurrences in the PBDB are also sometimes attributed to a reference that actually contains no data but that instead cites as its data source the PBDB or some other archive that we did not access. A more common source of discrepancy involves the injection of facts and interpretations by humans during data entry. Notably, approximately 50% of the ages assigned to fossil occurrences in the human database are not actually mentioned in the cited reference (Supplementary Fig. 6). Although problematic in some senses, this is well justified scientifically. The stated age for an occurrence in a document is often not the best available age, and the PBDB has no capacity to dynamically assign ages based on all evidence. Humans attempt to account for these limitations by entering what they determine, on the basis of other evidence, to be the best age for a fossil occurrence in a document. PDD replicated aspects of this behavior by inferring across all documents the most precise and recently published age for a given geological unit and location, but this is not sufficient to cover the full range of sources consulted by humans. Thus, a disproportionate number of the occurrences extracted by PDD have a temporal resolution (e.g., period-level) that results in their exclusion from the macroevolutionary quantities shown in Fig. 1. Including occurrences with low temporal resolution causes the absolute values of the human- and machine-generated diversity curves to converge (Supplementary Fig. 7).

Errors and limitations in the current PDD system also account for divergence in results (Fig.1). For example, OCR-related document processing failures, often involving tables, are among the leading causes of omissions by PDD (Supplementary Table 6). The current version of PDD also has design elements that cause some facts to be omitted. For example, PDD places great importance on formal geologic units, which means that no fossil occurrences are recognized in references that do not have well defined geologic units. Because this situation is more prevalent in recent time intervals, the lower total diversity recovered by PDD towards the recent (Fig. 1) is attributable to this design decision. Omissions also occur when a fact is correctly extracted by PDD, but with a probability < 0.95. This type of confidence-related error can typically be overcome by defining new features or rules.

The results from the ODS experiment demonstrate that our system performs comparably to humans in many complex data extraction and inference tasks and that macroevolutionary patterns are similarly expressed in both databases. This is an important result that demonstrates the reproducibility of key macroevolutionary results and that addresses several long-standing challenges in computer science. However, it is also the case that macroevolutionary quantities, which are based on large numbers of taxa, are robust to random errors introduced at the level of individual facts (32-34). Thus, the macroevolutionary results (Fig. 1) could be interpreted as evidence for the presence of a strong signal in the palaeontological literature that is readily recovered. The narrow distribution of range offsets on a per-genus basis (Fig. 2), however, suggests that PDD's precision is high even at the scale of individual facts.

**Training Data Requirements.** We used the human-constructed PBDB as both a source of training data and as a benchmark for evaluation. Therefore, an obvious question is, how big would the human database have to be in order for there to be sufficient training data to obtain a high quality result?

To assess the effect of training data volume on the quality of PDD, we randomly sampled the human database to produce a series of smaller databases. We then re-ran the entire PDD system in exactly the same way, but using only the subsampled data for training purposes. As expected, both the amount of data extracted by PDD (with a probability ≥ 0.95) and the accuracy of those data, summarized as the Spearman rank-order correlation between first differences in genus-level diversity (Fig. 1c), increases with the amount of training data. However, rather little training data is required in order to achieve a similarly high-quality result (Fig. 3). If the PBDB were populated with just 2% of the total number of references entered by humans over nearly two decades, there would be sufficient training data to obtain a comparable result.

**Whole Document Set (WDS).** Scaling PDD up to extract data from every relevant published document poses little technical challenge (35) and would offer a statistical advantage that could improve the overall quality of our system. However, access to the scientific literature for the purpose of automated text and data mining is currently limited (36). Thus, PDD's entire document set now consists of only 294,463 documents (Supplementary Table 8). Notably for this study, many of these documents were obtained from the open-access Biodiversity Heritage Library, which contains a large number of valuable but older and taxonomically-focused publications.

Despite limitations on our ability to access much of the relevant palaeontological literature, the PDD-generated Phanerozoic diversity curve for the WDS (Fig. 4) yields a face-value empirical genus diversity history that is congruent with classical estimates (3,4). First differences in Phanerozoic diversity extracted from the WDS are also positively correlated with first differences in diversity for the whole PBDB (Table 1). Genus-level rates of extinction and origination are also similar in both compilations (for first differences, $p < 0.0004$). The diversity histories of major groups of organisms comprising this total diversity are also positively correlated (Table 1), even though fewer than 25% of the references in the PBDB were read and processed by PDD (a total of 22,250 valid genera with resolved stratigraphic ranges are common to both compilations).

**Discussion**

The results of our validation study have three important implications. First, we have demonstrated that our machine reading system is capable of building a structured database from the heterogeneous scientific literature with quality that is comparable to, and in some cases possibly even exceeding, that produced by human readers (at least in the dimensions addressed here). This is notable because current benchmarks in machine reading and knowledge base construction, such as the Text Analysis Conference Knowledge Base Population competition, achieve less than 50% accuracy (albeit in the broader domain of general web text). Second, we have tested at a large scale the reproducibility of the PBDB, and in so doing we have identified sources of error and inconsistency that have a bearing on the use of the database. However, we have also shown that key macroevolutionary results are robust to these types of errors. Third, we have shown more broadly that literature-based macroevolutionary patterns are similarly expressed even when they derive from different bodies of literature. This indicates that the palaeontological literature, and presumably the underlying fossil record that it has sampled, contains a strong macroevolutionary signal that is readily recovered. This does not mean that our understanding of the global fossil record is uniformly complete taxonomically or in time and space (Supplementary Fig. 8), that our understanding of the true history of global biodiversity is accurate (12,13, 37,38), or even that the literature contains accurate data for every clade (e.g., 34).

The ability to expand existing databases and to more rapidly create new high quality synthetic data resources is a notable advance in the methodological toolkit of scientists. However, a much greater advantage of our approach is that the type of database that it produces is fundamentally different from manually populated databases. In the probabilistic database (28) produced by PDD, every fact is associated with an estimated probability of being correct and each fact remains tightly coupled to its original context. Thus, the quality of the entire database can be improved systematically whenever feedback is given on any one component or when additional rules or data is added to the system. More importantly, PDD's data acquisition process is based on the visual and textual analysis of entire documents. Our system is, therefore, able to recognize and extract data that are not currently part of a database but that are contextually related.

For example, the illustration of specimens is central to biological systematics and there are millions of biological illustrations in the WDS. Body size, a fundamental property of organisms that determines many aspects of their ecology (e.g., 39), is one of the morphological attributes readily conveyed by illustrations and their associated text. Several studies have examined the evolution of body size in individual lineages (e.g., 6), but, similar to the PBDB, all efforts to manually compile body size data cover only a small portion of the literature and yield monolithic databases that are difficult to assess and extend with new data.

To test the ability of our machine reading and learning system to incorporate data in illustrations, we extended PDD to identify images of biological specimens, locate and measure their major and minor axes, and read associated figure labels, captions, and text in order to determine magnification, the portion of the organism being imaged, and taxonomy (see Supplementary Information). The PDD-estimated body sizes for classified brachiopod genera are congruent with body sizes estimated for those same genera by the manual measurement of images (Fig. 5). Leveraging PDD's capacity to quantitatively analyze the entire body of published biological illustrations, in the context of their full textual descriptions, will enable new approaches to biological systematics and brings within reach questions that require a combination of morphological, geologic, and taxonomic data. Before PDD can be deployed to leverage this new capability, however, the current barriers to automated access and processing of published scientific documents must be overcome.

Although we have focused here on validating PDD and on testing the robustness of literature-derived macroevolutionary patterns in an widely used human-constructed database, our approach is built upon on a general machine reading and learning system (18) that can be readily adapted for many different domain-specific data extraction and inference tasks. We have shown that voluminous training data are not required to achieve high quality results. Thus, many questions that have been posed before, but that have been deemed too difficult to address without prohibitively time consuming data compilation efforts, are now within reach. Perhaps more importantly, our new approach to data synthesis yields a fundamentally different type of probabilistic database that remains tightly coupled to primary sources, that improves with the addition of new information, and that is capable of integrating complex data in ways that are likely to stimulate entirely new modes of inquiry.

**Methods**

Methods and any associated references are available in the online version of the paper.

**ACKNOWLEDGMENTS.** We thank M. Foote for constructive feedback and N.A. Heim and J.L. Payne for providing body size data. Work partially supported by NSF EarthCube award ACI-1343760 and NSF CAREER IIS-1353606. We also acknowledge the support of the Defense Advanced Research Projects Agency XDATA Program under No. FA8750-12-2-0335 and DEFT Program under No. FA8750-13-2-0039, the Office of Naval Research No. N000141210041 and No. N000141310129, Sloan Research Fellowship, American Family Insurance, Google, and Toshiba. Any opinions, findings, and conclusion or recommendations expressed in this material are those of the authors and do not necessarily reflect the view of these organizations or the US government. This is Paleobiology Database publication 2XX.

**Online Methods**

**System:** Features that relate facts in PDD are encoded in a relational database. These features derive from two sources: a set of functions written in the DeepDive framework and a set of existing tools developed by other researchers, including Tesseract and Cuneiform for text, Abbyy Fine Reader for tables, and StanfordCoreNLP for linguistic context. The list of features and rules used in this version of PDD are summarized in Supplementary Tables 1 and 2.

After extracting features in documents, the next step is to generate a factor graph (Supplementary Fig. 3), which is a compact way of specifying exponential family probability models (19, 40). The factor graph is defined by a hypergraph (*V*, *E*) where *V* is a set of random variables and $E \subseteq 2^V$ define groups of variables (factors) that are correlated. In addition, each random variable is associated with a domain (for simplicity, consider a Boolean random variable). Each factor (edge) $e = (v_1, \ldots, v_k)$ is associated with a scalar function called a potential (weight) $\varphi_e : \{0, 1\}^k \rightarrow R$. For example, the tuple (Tsingyuan Fm, Namurian) corresponds to a random variable, which assumes the value 1 if true. To specify a correlation, for example, if (Tsingyuan Fm, Carboniferous) is true, then it is likely that (Tsingyuan Fm, Namurian) is also true, a factor can be encoded to relate the two variables. This factor is only a statistical implication; PDD will estimate the strength of this implication on data.

The factor graph in PDD can be conceived of as existing in three layers (Supplementary Fig. 3). The first layer corresponds to the set of entities detected as individual mentions in documents. The second layer corresponds to a set of relation candidates between mentions, and the third layer corresponds to a set of relation candidates between distinct entities. One can think of the second layer as a per document layer and the third layer as the "aggregation" across all documents. Conceptualization of these layers is useful for software engineering reasons, but the statistical apparatus uses information from all layers simultaneously at the inference and learning stages.

Given a factor graph generated by feature extraction, PDD next learns the weight for each factor and then runs inference tasks to estimate the probability of each random variable. One key challenge of machine reading approaches is how to generate training data (i.e., a set of random variables that have been assessed for accuracy and that contain positive and/or negative examples). Traditional approaches include human expert annotation of results and crowd-sourcing (41). The human-constructed PBDB allows PDD to make extensive use of a generalization of Hearst patterns called distant supervision (42,43). This approach to training has considerable potential in the natural sciences because even simple lists of facts, such as the location and general geological age of rock formations, can be used in distant supervision to improve the quality of data extractions and more complex inferences.

Factor graphs are a convenient way to define random variables and their correlations, but they can be large. In PDD, the factor graph contains more than 200 million random variables and 300 million factors with 12 million distinct weights (Supplementary Table 9). PDD uses recent research in both theory (23, 24) and systems (25) to address this computational challenge. Further details are given the Supplementary Information.

**Documents.** The serial publications used in the ODS and WDS are provide din Supplementary Tables 1 and 8. Some of the serials in the top-50 PBDB sources were not accessible to us online. We were also not able to able to recover all references in the PBDB, due primarily to incomplete bibliographic information in the human database (Supplementary Tables 10, 11) and OCR and NLP document processing failures (see Assessment, below). To match retrieved documents to specific PBDB references we first used the TokenSet Cosine similarity approach (44) and then created an Amazon Mechanical Turk job, in which 64 distinct human workers combined for 30,182 evaluations of the matches. To obtain the WDS, we extended the ODS to include all available documents in the top-50 serials in the PBDB; we also included the whole Biodiversity Heritage Library.

**Features.** All PDD feature extraction tasks that use existing tools were run on Condor and the Open Science Grid (OSG). Ghostscript was run to convert each document into a set of png images. Next, OCR tools were executed. Each tool was permitted to run for 24 hours on a document before timeout occurred; a failed document was re-deployed on the OSG up to 10 times before being removed from the set. Document failures were caused by kernels older than 2006 and incompatible software on individual OSG machines, as well as document-specific software bugs, such as segmentation faults in Cuneiform caused by unusual document formatting. All tools had a failure rate of less than 8%, but these errors are orthogonal to our work; future improvements to these tools will improve PDD.

The WDS contains 23 times more documents than the ODS, and the number of variables extracted from them scales approximately linearly. The number of distinct features is, however, only 13 times greater because features can be shared across documents (Supplementary Table 12). Distinct taxa are only 10 times more numerous in the WDS because many taxa are referred to in more than one document. The number of occurrences is only six times greater in the WDS, reflecting the fact that most of the additional documents we were able to access are taxonomically-focused and do not contain fossil occurrence data; some documents also derive from serials, such as USGS Open-File Reports, that are interdisciplinary and have only a minority of documents relevant to palaeontology.

**Extensions.** We extended PDD to include data extraction from German and Chinese language documents. The named entity recognition component of PDD has dictionary-based features and NLP-based features. Relevant language-specific dictionaries were built manually and from external sources such as geonames.org. For NLP-based features, the Stanford CoreNLP provides models for Chinese and German. For document layout-based features, there is no change in function with language.

We also extended PDD to extract body size from biological illustrations, which requires processing images, linking image part labels to captions, and mapping captions to text in order to extract all of the necessary information (Supplementary Fig. 9, 10). Explanation of tools and methods used for joint image-text analysis is presented in the Supplementary Information.

**Assessment.** The ODS was randomly and evenly split into a training set and a testing set. Fifty documents in the testing set were then randomly sampled for assessment by human annotators, primarily graduate students in the Dept. of Geoscience at UW-Madison. Assessments included taxonomic, stratigraphic, chronologic, and geographic tuples. PDD achieves ≥ 92% human-estimated accuracy in all relations (Supplementary Table 13), which is close to the 95% confidence threshold specified for data output.

The number of facts recovered vs. the number of facts contained in a document (i.e., recall) is more difficult to assess than the precision of extracted data. Because each extracted relationship consists of a paired object and subject (e.g., the object "formation" contains a subject "taxon"), one basic measure of recall is the fraction of all subjects in the PBDB that PDD also recovered. This estimate of recall ranges from 21% to 69%, depending on relation (Supplementary Table 13). For the lowest recall relations, we randomly sampled 10 documents in order to compare the PBDB and PDD. We did so for a combination of three binary relations (taxon,formation)(formation,temporal)(formation,location). When summarizing this 4-part tuple by projecting these relationships to taxon, approximately 18% of PDDs extractions also appear in PBDB and 11% of PBDB extractions also appear in PDD. This implies that both PDD and PBDB make recall errors, but that both systems also have high precision. Further examination of PDD recall errors (Supplementary Table 6) shows that they can be attributed to OCR-related failures (56%), table recognition problems (29%), and lack of context features required to address the full range of complicated expressions in the literature (15%). All of these errors correspond to interesting and open-problems for computer science. The first two are related to data acquisition (i.e., how to correctly recognize the structure and content of a given document), and the latter is an important natural language inference problem (i.e., how to extract relations by taking advantage of information in the whole document). Continued work in these areas will further improve the PDD system, which we have shown is already capable of meeting, and in some cases exceeding, human standards in its ability to produce a synthetic database resource with proven scientific value. For additional technical validation of the system, including an explanation of the calibration of probabilities in the database (Supplementary Figs. 11, 12) and the impact of including rich features on overall system quality (Supplementary Figs. 13, 14), see the Supplementary Information.

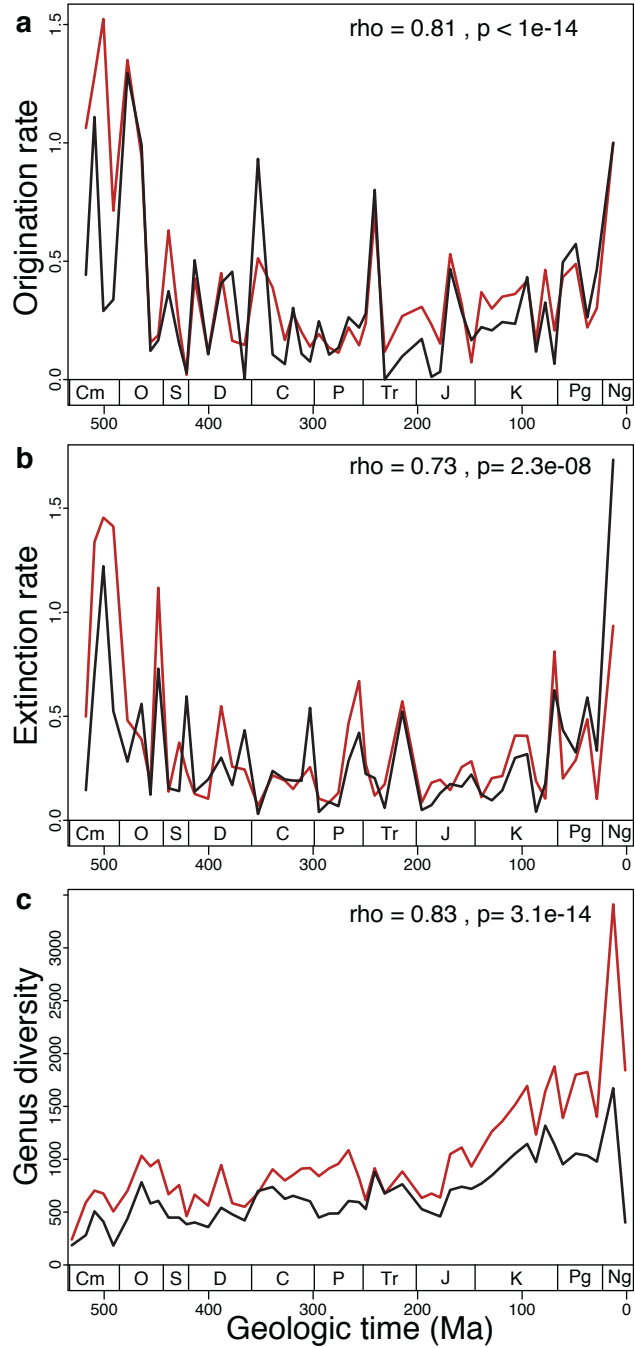

**Figure 1.** Machine- and human-generated macroevolutionary results for the overlapping document set. Human-generated in red, machine-generated in black. Spearman rank order correlations for first differences shown. (**a**) Per capita, per interval origination rates (29). (**b**) Per capita, per interval extinction rates. (**c**) Total range-through diversity.

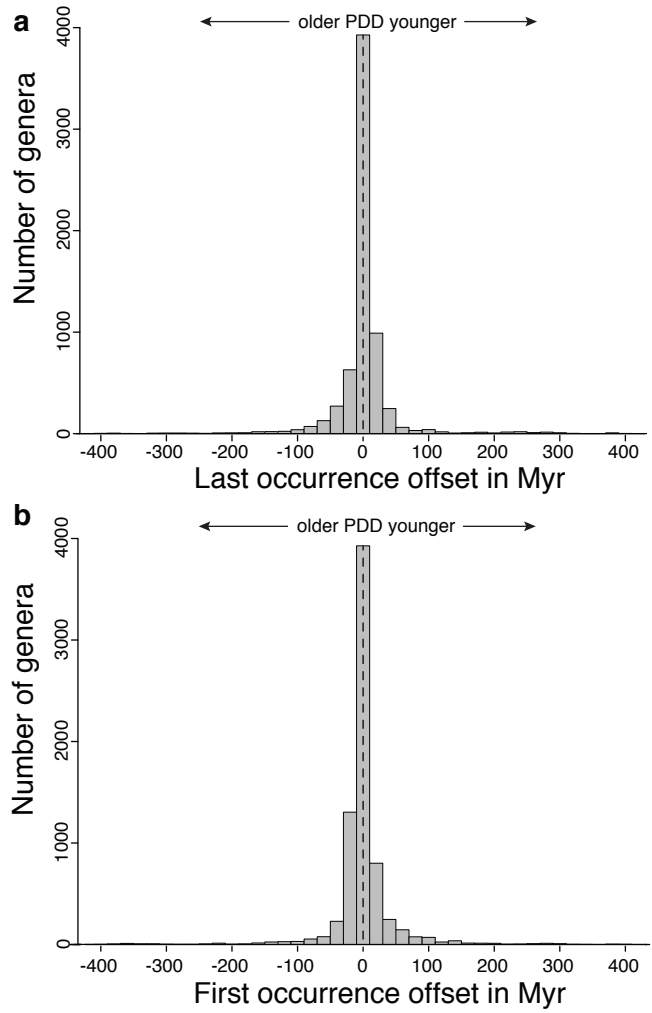

**Figure 2**. Difference in genus range end points for 6,708 genera common to the PBDB and PDD. (**a**) Last occurrence differences. Median is 0 Myr, mean is +1.7 Myr. (**b**) First occurrence offset. Median is 0 Myr, mean is -0.3 Myr.

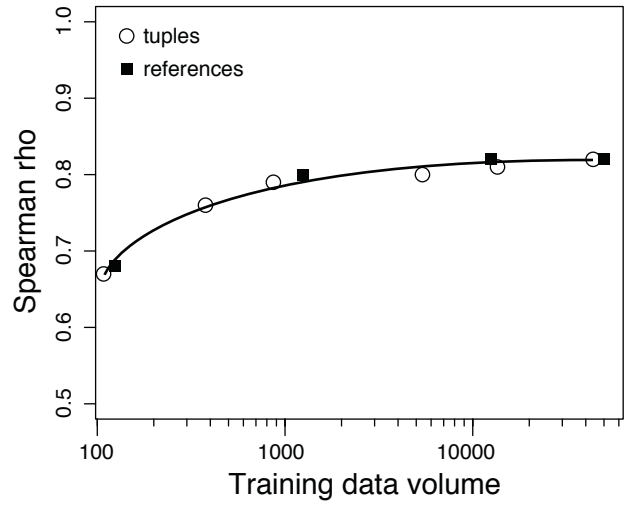

**Figure. 3**. Effect of changing PBDB training database size on PDD quality. Spearman rho is correlation between human- and machine- generated time series of diversity, as in Fig. 1c.

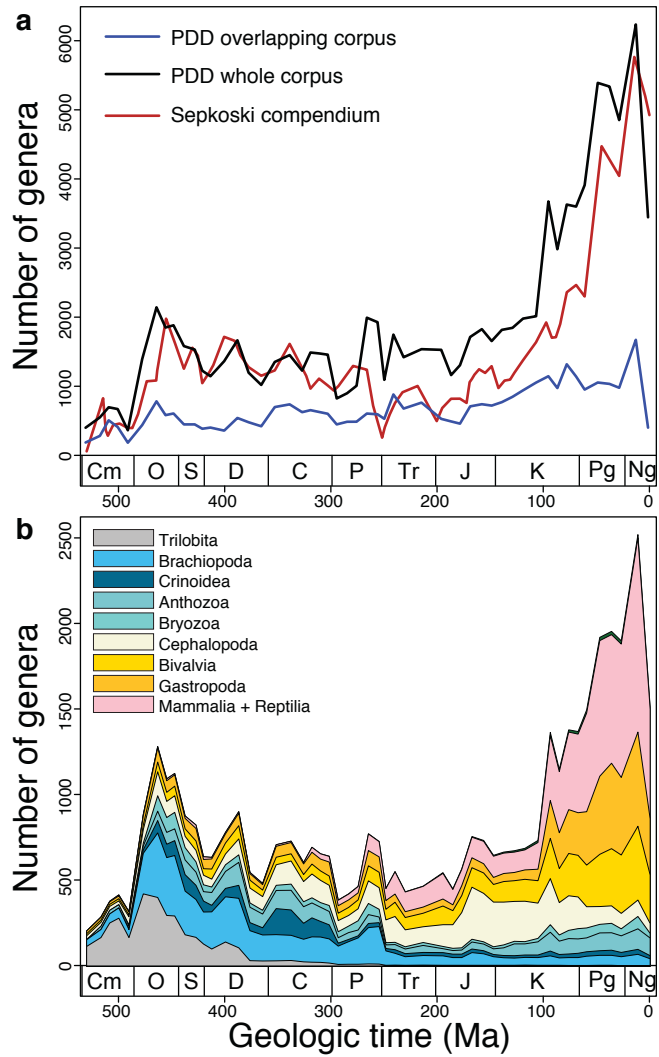

**Figure 4.** Genus-level diversity generated by PDD for the whole document set. (**a**) Total genus diversity calculated as in Fig. 1. For comparison, Sepkoski's genus-level diversity curve (3,4) is plotted using his stage-level timescale. (**b**), Diversity partitioned by genera resolved to select classes by PDD.

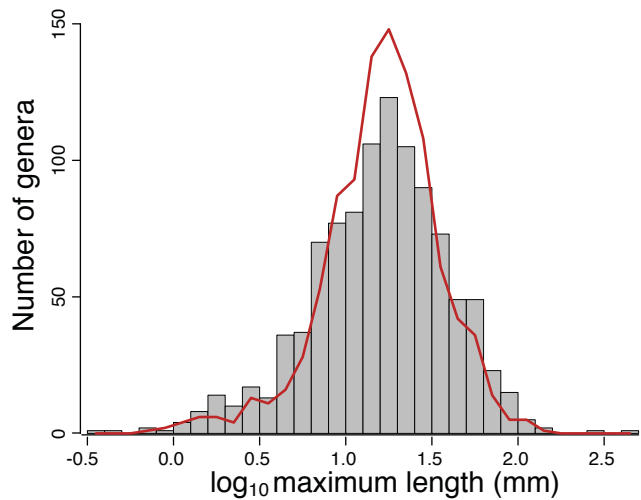

**Figure 5.** Frequency distributions of paired estimates of body size for 1,014 brachiopod genera. PDD, gray bars; human estimate, red line. Distributions not significantly different according to paired Mann-Whitney U-test (p = 0.18) and Kruskal-Wallis test (p = 0.64) .

**Table 1.** Genus-level diversity in the whole document set and the entire PBDB. Spearman rank-order correlation coefficients and p-values for detrended diversity time series (from Fig. 4b) shown.

| Taxonomic group | Spearman rho | P-value |
| --- | --- | --- |
| All genera | 0.72 | 3.6x10 |
| Bivalvia | 0.67 | 6.2x10 |
| Bryozoa | 0.64 | 3.6x10 |
| Gastropoda | 0.59 | 5.3x10 |
| Anthozoa | 0.53 | 6.6x10 |
| Brachiopoda | 0.52 | 0.0001 |
| Reptilia | 0.50 | 0.0002 |
| Trilobita | 0.49 | 0.0003 |
| Cephalopoda | 0.41 | 0.003 |
| Mammalia | 0.40 | 0.004 |
| Crinoidea | 0.39 | 0.004 |

# SUPPLEMENTARY INFORMATION

# A machine-compiled macroevolutionary history of Phanerozoic life

Shanan E. Peters [*], Ce Zhang [†], Miron Livny [†], and Christopher Ré [‡]

[*] Department of Geoscience, University of Wisconsin-Madison, Madison, WI, 53706 USA
[†] Department of Computer Science, University of Wisconsin-Madison, Madison, WI, 53706 USA
[‡] Department of Computer Science, Stanford University, Stanford, CA 94305 USA

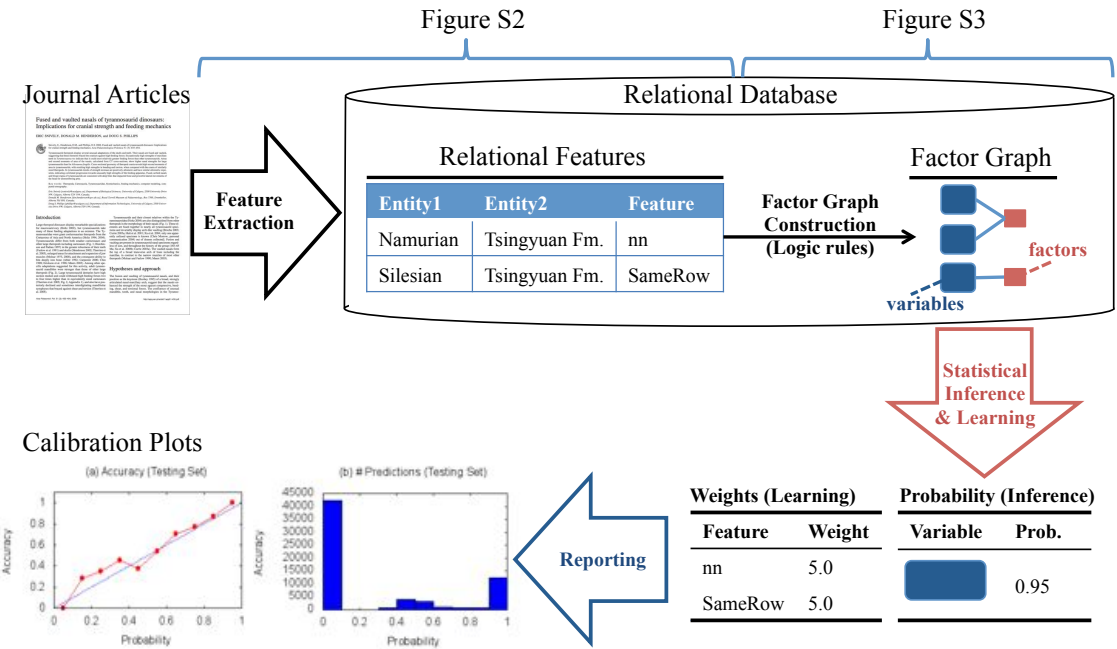

**Supplementary Figure 1.** Schematic representation of the PDD workflow.

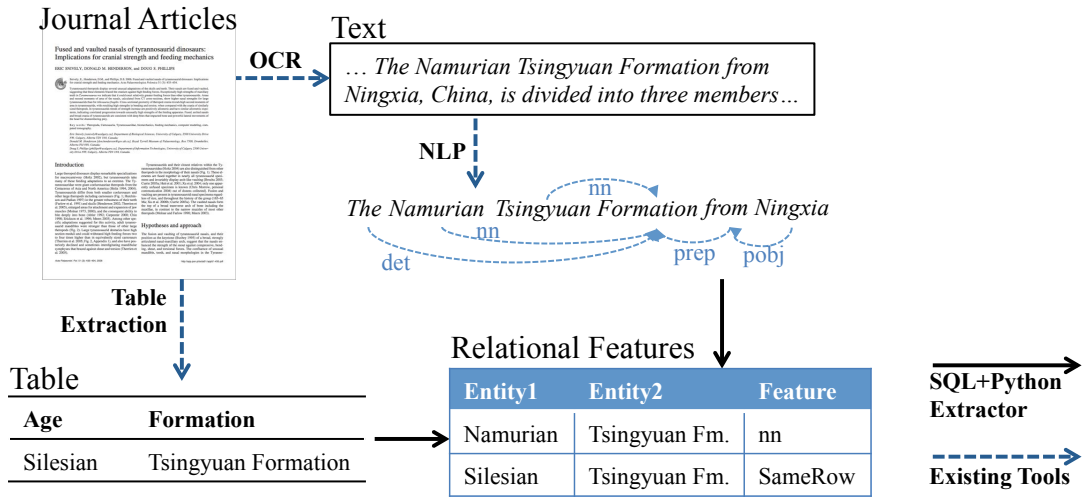

**Supplementary Figure 2.** Overview of PDD feature extraction. Text, tables, and images in an original document are parsed (e.g., by table position extraction or natural language). Two or more entities and the specific properties in the document (i.e., features) that relate them are expressed as a row in a database.

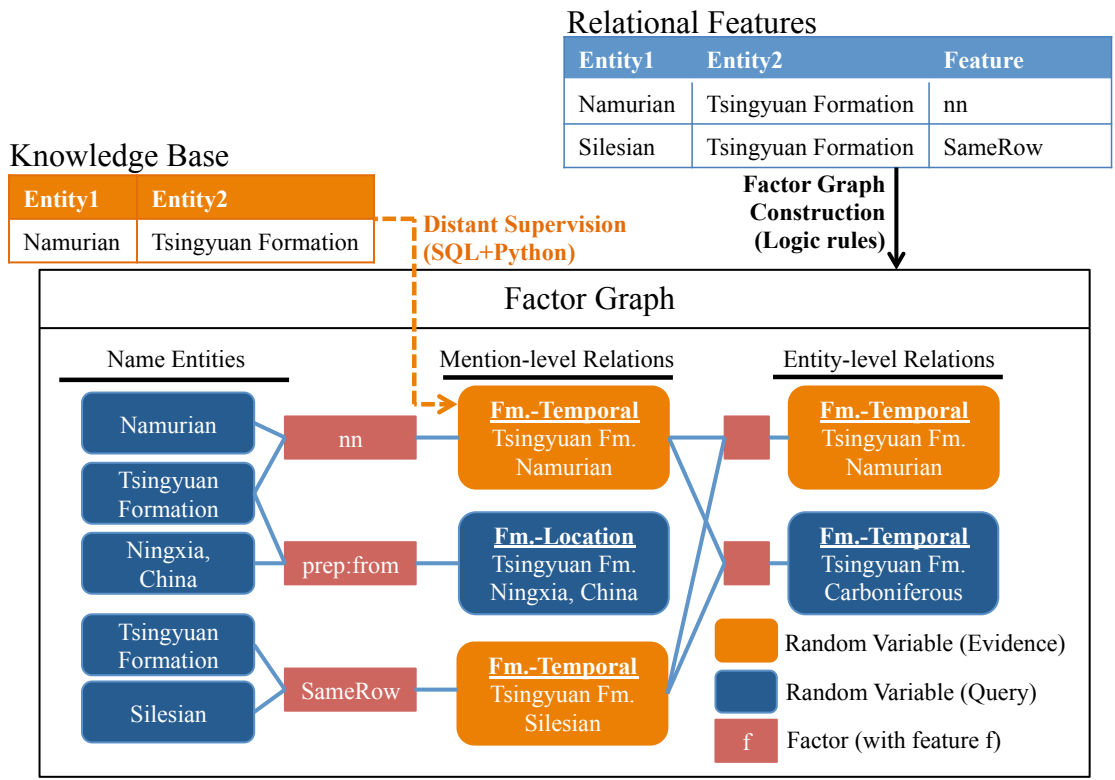

**Supplementary Figure 3.** Overview of factor graph component of PDD. Existing knowledge bases, such as data in the PBDB, are used to assess mention-level relations during distant supervision. Variables assessed for accuracy become evidence variables for statistical inference and learning steps.

| taxon | rank | status | taxon | rank | in ref | not in ref | incorrect | ? |
|---|---|---|---|---|---|---|---|---|
| Bellerophon carbonarius | species | belongs to | Euphemites | genus | | ■ | | |
| Bellerophon urii | species | belongs to | Euphemites | genus | | | ■ | |
| Bucaniopsis hibernicus | species | belongs to | Retispira | genus | | ■ | | |
| Euphemus konincki | species | belongs to | Euphemites | genus | ■ | | | |

**Supplementary Figure 4.** Screen shot of web user interface used in blind experiment conducted by 7 human annotators. A unique link and instructions to complete the form were emailed to each participant. The wording of the instructions was as follows:

1. "in ref" means you can find this *exact* fact in the document somewhere.

2. "not in ref" means you can't find the exact fact in the document anywhere (can include typos).

3. "incorrect" means it is an incorrect fact (e.g., wrong assignment/relationship, etc.).

4. "?" means you don't understand the fact in relation to document.

Simply clicking on the box selects it for you. You can change it etc. as you go along. Once you are done, you can go to another ref by clicking on bottom. You can come back to the ref and inspect it to make sure it looks good, change things.

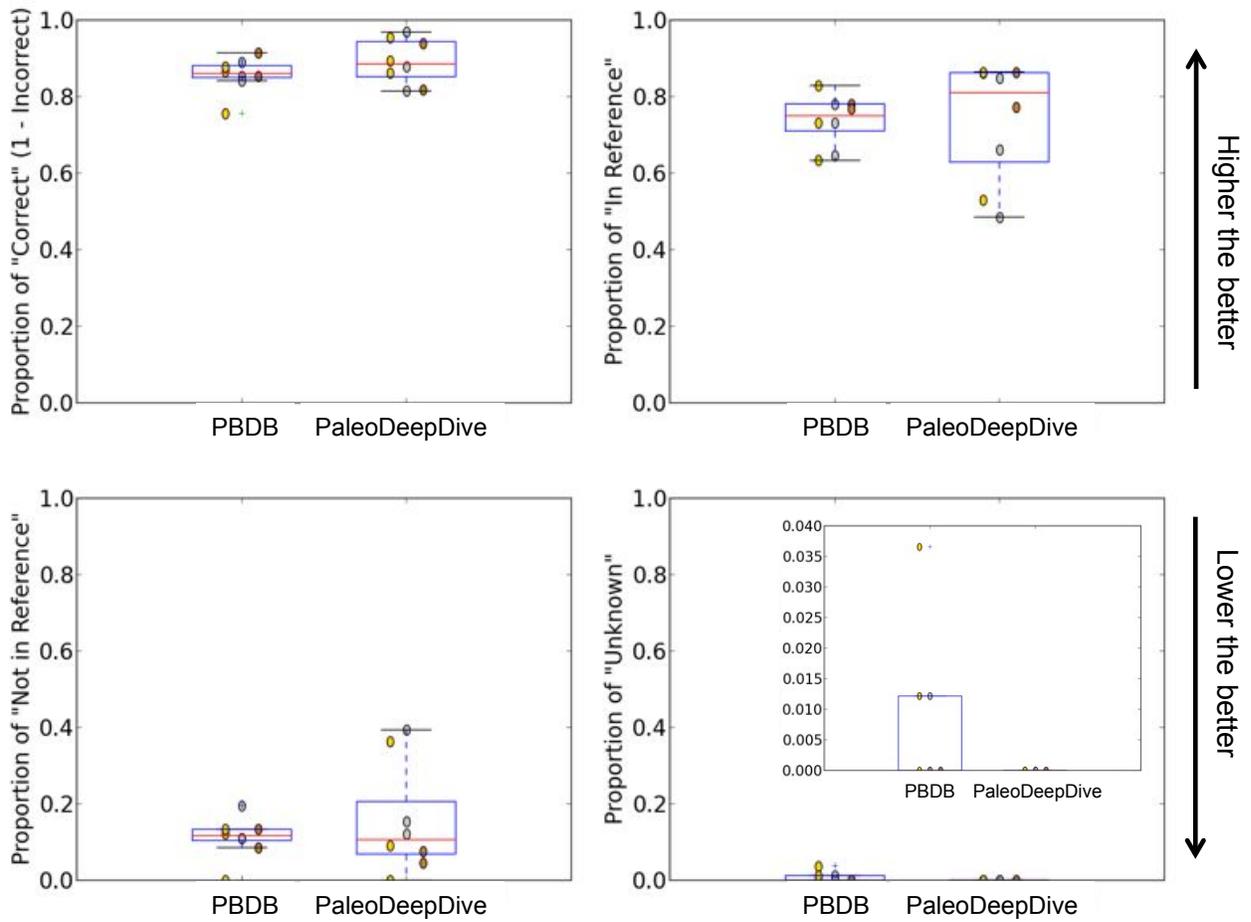

**Supplementary Figure 5.** Summary of results of annotation experiment of PDD and PBDB taxonomic extractions. Yellow, annotators with heavy PBDB governance involvement; blue, past governance involvement; red, graduate students.

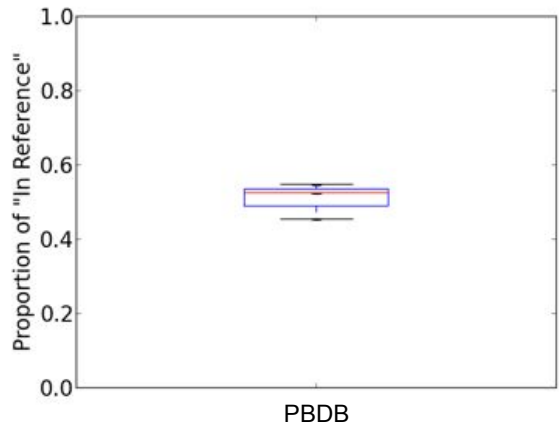

**Supplementary Figure 6.** Summary of results of annotation experiment of occurrence data, or (taxon, geologic unit, temporal interval) tuples in human-constructed PBDB. Results are for 3 volunteers, one from each of groups in Supplementary Figure 4.

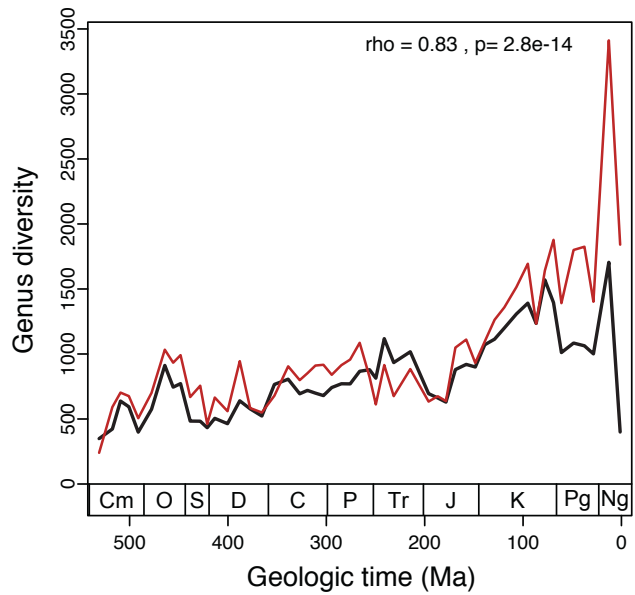

**Supplementary Figure 7.** PDD genus-level diversity (black curve) calculated using occurrences with period level or finer temporal resolution, as opposed to epoch or finer temporal resolution used in Fig. 1. The red curve shows PBDB data and is identical to the red curve in Fig. 1c.

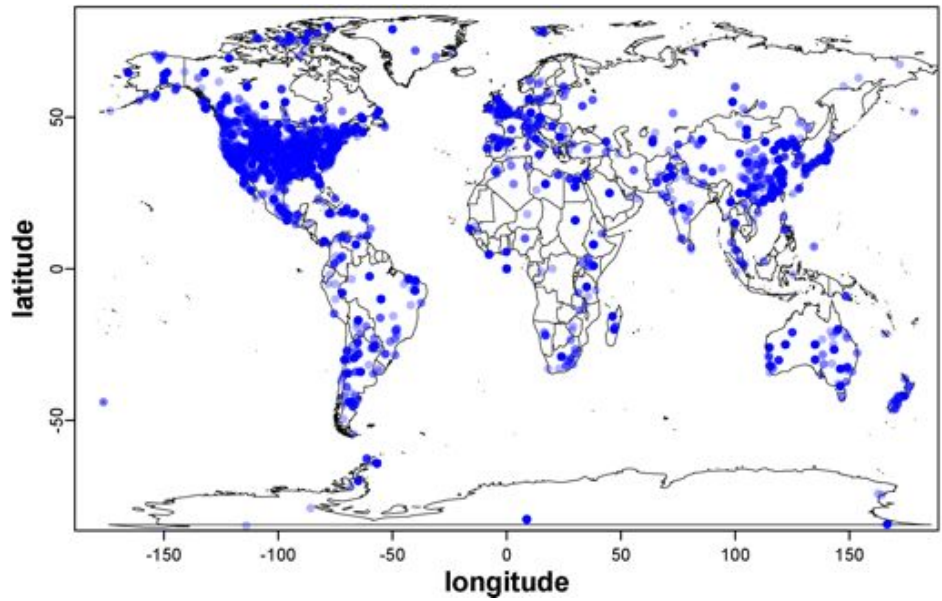
(a) Overlapping Corpus

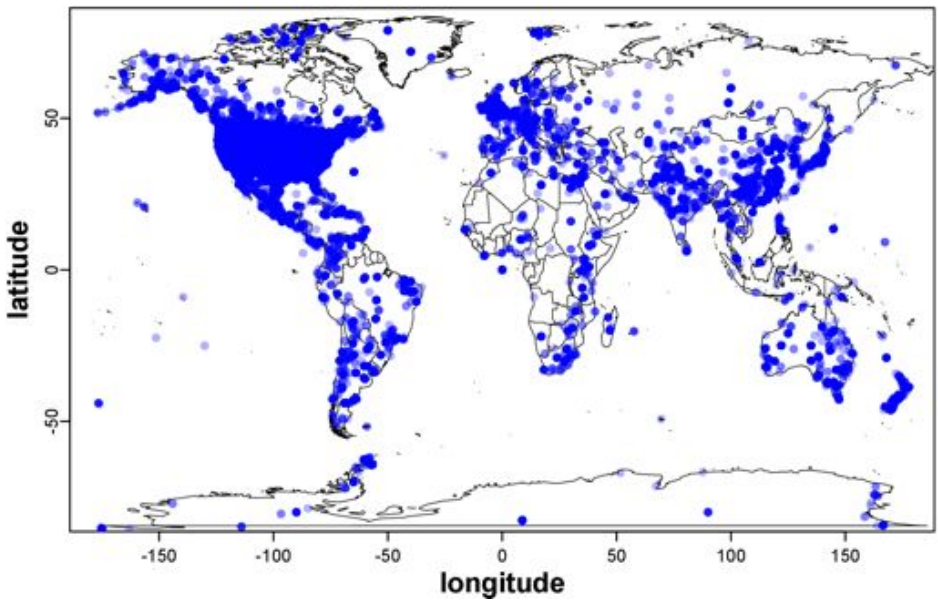
(b) Whole Corpus

**Supplementary Figure 8.** Geographic distribution of PDD-generated database. Top, location of occurrences in overlapping document set (ODS). Bottom, location of occurrences in whole document set (WDS).

| Layer | Features |
|---|---|
| Name Entities | Dictionary (English dictionary, GeoNames, PaleoDB, Species2000, Microstrat, MySQL stop words) |
| | Part-of-speech tag from StanfordCoreNLP |
| | Name-entity tag from StanfordCoreNLP |
| | Name entity mentions in the same sentences (paragraphs, or documents) |
| Mention-level Relations | Word sequence between name entities |
| | Dependency path between name entities |
| | Name-entity tag from StanfordCoreNLP |
| | Table caption-content association |
| | Table cell-header association |
| | Section headers (for Taxonomy) |
| Entity-level Relations | Temporal interval containment (e.g., Namurian $\subseteq$ Carboniferous) |
| | Location containment (e.g., Ningxia, China $\subseteq$ China) |
| | One formation does not likely span $>$ 200 million years |

**Supplementary Table 1.** List of features and rules used in the current verison of PDD. Finding the right simple features and rules can be difficult. The PDD system is designed to operate in an iterative fashion, with error analysis occurring after each round of feature and rule definition.

| Relation | Tuple in Knowledge | Positive Examples | Negative Examples |
|---|---|---|---|
| Taxonomy | (Taxon, Taxon) $(t_1, t_2)$ | $\{(t_1, t_2)\}$ | $\{(t_1, t_2') : t_2' \neq t_2\}$ |
| Formation | (Taxon, Formation) $(t, f)$ | $\{(t, f)\}$ | Positive examples of other relations |
| Formation-Temporal (Mention) | (Formation,Interval) $(t, i)$ | $\{(t, i') : intersect(i, i')\}$ | $\{(t, i') : \neg intersect(i, i')\}$ |
| Formation-Temporal (Entity) | (Formation,Interval) $(t, i)$ | $\{(t, i') : intersect(i, i') \wedge \neg contain(i', i)\}$ | $\{(t, i') : \neg intersect(i, i')\}$ |
| Formation-Location (Mention) | (Formation,Location) $(t, l)$ | $\{(t, l') : intersect(l, l')\}$ | $\{(t, l') : \neg intersect(l, l')\}$ |
| Formation-Location (Entity) | (Formation,Location) $(t, l)$ | $\{(t, l') : intersect(l, l') \wedge \neg contain(l', l)\}$ | $\{(t, l') : \neg intersect(l, l')\}$ |

**Supplementary Table 2.** List of distant supervision rules used in PDD. Function $contain(x, y)$ and $intersect(x, y)$ return True if the interval (or locations) $x$ contains or intersects with $y$.

| Journal Name | PBDB | PDD Overlapping Set | Coverage |
|---|---|---|---|
| Journal of Paleontology | 2,667 | 2,534 | 95% |
| Journal of Vertebrate Paleontology | 1,909 | 1,292 | 68% |
| Palaeontology | 879 | 748 | 85% |
| Paleontological Journal | 849 | 0 | 0% |
| American Museum Novitates | 513 | 433 | 84% |
| NULL | 509 | 0 | 0% |
| Acta Palaeontologica Polonica | 483 | 433 | 90% |
| Nature | 452 | 340 | 75% |
| Cretaceous Research | 424 | 421 | 99% |
| Gobios | 423 | 296 | 70% |
| Ameghiniana | 394 | 21 | 5% |
| Canadian Journal of Earth Sciences | 336 | 281 | 84% |
| Palaeogeography, Palaeoclimatology, Palaeoecology | 325 | 317 | 98% |
| Vertebrata PalAsiatica | 322 | 203 | 63% |
| Science | 309 | 184 | 60% |
| Bulletin of the American Museum of Natural History | 293 | 214 | 73% |
| Geological Magazine | 269 | 24 | 9% |
| Alcheringa | 268 | 0 | 0% |
| American Journal of Science | 257 | 53 | 21% |
| Palaeontologische Zeitschrift | 241 | 0 | 0% |
| Journal of Mammalogy | 234 | 147 | 63% |
| Acta Palaeontologica Sinica | 232 | 3 | 1% |
| United States Geological Survey Professional Paper | 231 | 156 | 68% |
| Zoological Journal of the Linnean Society | 203 | 200 | 99% |
| Contributions from the Museum of Paleontology, University of Michigan | 195 | 174 | 89% |
| Palaeontographica Abteilung A | 194 | 0 | 0% |
| Facies | 187 | 0 | 0% |
| Lethaia | 183 | 178 | 97% |
| Quarterly Journal of the Geological Society of London | 180 | 122 | 68% |
| Zootaxa | 180 | 0 | 0% |
| Palaios | 174 | 164 | 94% |
| Annals of Carnegie Museum | 172 | 25 | 15% |
| Proceedings of the United States National Museum | 149 | 0 | 0% |
| Neues Jahrbuch fr Geologie und Paleontologie, Abhandlungen | 147 | 0 | 0% |
| Review of Palaeobotany and Palynology | 147 | 146 | 99% |
| American Journal of Botany | 147 | 87 | 59% |
| Proceedings of the Academy of Natural Sciences of Philadelphia | 142 | 40 | 28% |
| Journal of Human Evolution | 135 | 122 | 90% |
| Proceedings of the National Academy of Sciences | 133 | 51 | 38% |
| Journal of Systematic Palaeontology | 132 | 27 | 20% |
| Geodiversitas | 131 | 0 | 0% |
| Acta Geologica Sinica | 130 | 78 | 60% |
| Bulletins of American Paleontology | 129 | 0 | 0% |
| Bulletin de la Societe Geologique de France | 122 | 0 | 0% |
| Palontologische Zeitschrift | 115 | 0 | 0% |
| Rivista Italiana di Paleontologia e Stratigrafia | 115 | 0 | 0% |
| Psyche | 111 | 1 | 1% |
| Annals of the South African Museum | 104 | 0 | 0% |
| Tulane Studies in Geology and Paleontology | 103 | 0 | 0% |
| Paleontological Research | 102 | 92 | 90% |
| Other Sources | 30,851 | 2,175 | 7% |
| **Total** | **47,632** | **11,782** | **25%** |

**Supplementary Table 3.** Distribution of documents in the overlapping document set. "NULL" corresponds to a NULL title document type field in the PBDB.

| Taxon Name | Rank | Not Found on Google (Error Candidate) |
|---|---|---|
| Cirquella espinata | species | |
| Echinophyllia orpheensis | species | |
| Fenestella huascatayana | species | |
| Epigondolella primitia | species | |
| Palaeospheniscus gracilis. | species | |
| Pygurus carinatus | species | × |
| Arionellus tripunctatus | species | |
| Phacostylus amphistylus | species | |
| Circotheca multisulcatus | species | |
| Aulotortus praegaschei | species | |
| Leptaena demissa | species | |
| Xinjiangchelys laticentralis | species | |
| Conotreta lanensis | species | × |
| Martellia ichangensis | species | |
| Procavia antiqua | species | |
| Chermidae | family | |
| Monophyllus cubanus | species | |
| Gazella soemmeringi | species | |
| Pinna subspatulata | species | |
| Polacanthus faxi | species | × |
| Homotherium latidens | species | |
| Platanus primaeva | species | |
| Rhopalocanium satelles | species | |
| Cryptobairdia forakerensis | species | |
| Naiadites elongata | species | |
| Staurocephalus murchisoni | species | |
| Serpula anguinus | species | |
| Glycymeris angusticostata | species | |
| Eomunidopsis eutecta | species | |
| Actinocrinites gibsoni | species | |
| Zhelestes tes | species | × |
| Spinocyrtia ascendens | species | |
| Belemnopsis alexandri | species | |
| Agaricocrinus nodulosus | species | |
| Oreochromis shiranus | species | |
| Atrichornithidae | family | |
| Neltneria jaqueti | species | |
| Eurydice affinis | species | |
| Nummulites burdi | species | |
| Diacalymene marginata | species | |
| Scapteriscus didactylus | species | |
| Enhydriodon campanii | species | |
| Offneria nicoli | species | × |
| Propetrosia pristina | species | |
| Podocarpus campbelli | species | |
| Graffhamicrinus aristatus | species | |
| Productina sampsoni | species | |
| Bufina bicornuta | species | |
| Coccolithus staurion | species | |
| Ernanodon vas | species | × |

**Supplementary Table 4.** Error Analysis of Taxon Entity Extractions in PDD

| Reference No. | Genus | Correct | Extracted by PBDB |
|---|---|---|---|
| 28945 | Acrodenta | ✓ | |
| | Mastodonsaurus | ✓ | |
| | Mesodapedon | ✓ | |
| | Rhynchosaurus | ✓ | ✓ |
| | Scaphonyx | ✓ | |
| | Spirorbis | ✓ | |
| | Stenaulorhynchus | ✓ | |
| 34109 | | | |
| 28146 | | | |
| 38697 | Hazelia | ✓ | ✓ |
| | Leptomitus | ✓ | ✓ |
| 32675 | | | |
| 33994 | Gastropoda | | |
| | Heterostropha | ✓ | |
| | Mathilda | ✓ | |
| | Mollusca | ✓ | |
| | Stenoglossa | ✓ | |
| 27115 | | | |
| 41374 | Archaeopterodactyloidea | ✓ | |
| | Beipiaopterus | ✓ | |
| | Boreopteridae | ✓ | |
| | Boreopterus | ✓ | |
| | Eopteranodon | ✓ | |
| | Eosipterus | ✓ | |
| | Feilongus | ✓ | |
| | Gegepterus | ✓ | |
| | Moganopterus | ✓ | ✓ |
| | Ningchengopterus | ✓ | |
| | Ornithocheiroidea | ✓ | |
| | Zhenyuanopterus | ✓ | |
| 12054 | | | |
| 13061 | Bactrosaurus | ✓ | |
| | Dyoplosaurus | ✓ | |
| | Gorgosaurus | ✓ | |
| | Hypacrosaurus | ✓ | |
| | Mandschurosaurus | ✓ | ✓ |
| | Nodosauridae | ✓ | ✓ |
| | Tanius | ✓ | |
| Human Recall | | | 18% |

**Supplementary Table 5.** Error Analysis: PDD Extractions

| Reference No. | Genus | Correct | Extracted by PDD | Error Reason |
|---|---|---|---|---|
| 28945 | Rhynchosaurus | ✓ | ✓ | |
| 34109 | Austromola | ✓ | | Not enough context features |
| | Odontoceti | ✓ | | Not enough context features |
| 28146 | Cerapoda | ✓ | | Not enough context features |
| 38697 | Hazelia | ✓ | ✓ | |
| | Leptomitus | ✓ | ✓ | |
| | Protospongia | ✓ | | Not enough context features |
| 32675 | Tommotia | ✓ | | Not enough context features |
| 33994 | Anticonulus | ✓ | | |
| | Ataphrus | ✓ | | |
| | Austriacopsis | ✓ | | |
| | Discohelix | ✓ | | |
| | Emarginula | ✓ | | |
| | Eucyclidae | ✓ | | Table recognition failure |
| | Eucyclus | ✓ | | |
| | Guidonia | ✓ | | |
| | Neritopsis | ✓ | | |
| | Plectotrochus | ✓ | | |
| | Proacirsa | ✓ | | |
| | Pseudorhytidopilus | ✓ | | |
| 27115 | Astreptodictya | ✓ | | |
| | Athrophragma | ✓ | | |
| | Batostoma | ✓ | | |
| | Bryozoa | ✓ | | |
| | Bythopora | ✓ | | |
| | Calopora | ✓ | | |
| | Coeloclema | ✓ | | |
| | Constellaria | ✓ | | |
| | Contexta | ✓ | | |
| | Diploclema | ✓ | | |
| | Echinodermata | ✓ | | |
| | Graptodictya | ✓ | | OCR error |
| | Helopora | ✓ | | |
| | Nicholsonella | ✓ | | |
| | Ottoseetaxis | ✓ | | |
| | Pachydictya | ✓ | | |
| | Phylloporina | ✓ | | |
| | Porifera | ✓ | | |
| | Prasopora | ✓ | | |
| | Spongiostroma | ✓ | | |
| | Stictopora | ✓ | | |
| | Stictoporella | ✓ | | |
| | Trilobita | ✓ | | |
| 41374 | Moganopterus | ✓ | ✓ | |
| 12054 | Neosaurus | ✓ | | Not enough context features |
| 13061 | Mandschurosaurus | ✓ | ✓ | |
| | Nodosauridae | ✓ | ✓ | |
| PDD Recall | | | 11% | |

**Supplementary Table 6.** Error Analysis: PBDB Extractions

| Relation | PBDB | PDD | $p = 0.05$ |
|---|---|---|---|
| Taxonomy | 92% | 97% | 0 |
| Temporal | 89% | 96% | + |
| Location | 90% | 92% | 0 |
| Formation | 84% | 94% | + |

**Supplementary Table 7.** Comparison of Accuracies of PDD and PBDB. The column $p = 0.05$ is the significant test of one-tail Welch's $t$-test, where "+" means significant given the corresponding $p$-value, and "0" otherwise. The value 0.05 is picked by following the default setting of R.

| Journal Name | 1845-<br>-1959 | 1960-<br>-1969 | 1970-<br>-1979 | 1980-<br>-1989 | 1990-<br>-1999 | 2000-<br>-2009 | 2010-<br>-2013 | Total |
|---|---|---|---|---|---|---|---|---|
| American Journal of Science | 2489 | | 727 | 41 | | 245 | 138 | 3640 |
| American Midland Naturalist | 2893 | 1022 | 1149 | 989 | 852 | 842 | 189 | 7936 |
| American Museum Novitates | 1974 | 413 | 288 | 272 | 320 | 388 | 98 | 3753 |
| Annales de Palontologie | | | | | 29 | 206 | 73 | 308 |
| Annals of Carnegie Museum | | | | | | 82 | 38 | 120 |
| Bulletin of the American Museum of Natural History | 1318 | 93 | 105 | 72 | 52 | 196 | 65 | 1901 |
| Comptes Rendus Palevol | | | | | | 679 | 270 | 949 |
| Cretaceous Research | | | | 287 | 457 | 732 | 393 | 1869 |
| Geological Journal | 136 | 418 | 338 | 1116 | 680 | 662 | 423 | 3773 |
| Geological Society America Bulletin | 276 | 796 | | 788 | 1158 | 1089 | 486 | 4593 |
| Geology | | | 1177 | 2675 | 2990 | 3024 | 1261 | 11127 |
| Global and Planetary Change | | | | 20 | 469 | 1070 | 376 | 1935 |
| Gobios | | 13 | 442 | 1072 | 1294 | 753 | 167 | 3741 |
| International Geology Review | 87 | 1482 | 1780 | 1541 | 724 | 635 | 353 | 6602 |
| Journal of Asian Earth Sciences | | | | | 149 | 1162 | 1123 | 2434 |
| Journal of Geology | 5782 | 736 | 929 | 754 | 671 | 516 | 153 | 9541 |
| Journal of Human Evolution | | | 859 | 890 | 759 | 1067 | 597 | 4172 |
| Journal of Mammalogy | 3023 | 1633 | 1509 | 1452 | 1336 | 1506 | 438 | 10897 |
| Journal of Paleontology | 2552 | 1500 | 1438 | 1297 | 1172 | 2224 | 643 | 10826 |
| Journal of South American Earth Sciences | | | | 79 | 423 | 666 | 414 | 1582 |
| Journal of Systematic Palaeontology | | | | | | 113 | 110 | 223 |
| Journal of Vertebrate Paleontology | | | | 365 | 636 | 2152 | 934 | 4087 |
| Journal of the Geological Society | | | | | 329 | 946 | 346 | 1621 |
| Lethaia | | 104 | 830 | 978 | 992 | 738 | 371 | 4013 |
| Mammalian Species | | 1 | 122 | 224 | 284 | 216 | | 847 |
| Marine Micropaleontology | | | 85 | 262 | 469 | 646 | 156 | 1618 |
| Micropaleontology | 202 | 375 | 302 | 264 | 270 | 316 | | 1729 |
| New Zealand Journal of Geology and Geophysics | 121 | 733 | 730 | 519 | 484 | 403 | 115 | 3105 |
| PALAIOS | | | | 290 | 567 | 677 | 237 | 1771 |
| Palaeogeography, Palaeoclimatology, Palaeoecology | | 191 | 600 | 1108 | 1812 | 3221 | 1191 | 8123 |
| Palaeontology | 48 | 461 | 477 | 446 | 493 | 1470 | 560 | 3955 |
| Palaios | | | | | | 620 | 287 | 907 |
| Paleobiology | | | 184 | 422 | 337 | 866 | 260 | 2069 |
| Paleontological Research | | | | | | 192 | 88 | 280 |
| Palynology | | | 45 | 140 | 132 | 232 | 119 | 668 |
| Proc. of AASP | | | 79 | | | | | 79 |
| Proceedings of the Geologists' Association | 3514 | 430 | 415 | 416 | 404 | 394 | 273 | 5846 |
| Quarterly Journal of the Geological Society of London | 3063 | 177 | 19 | | | | | 3259 |
| Review of Palaeobotany and Palynology | | 241 | 427 | 705 | 1031 | 887 | 406 | 3697 |
| Revue de Micropaleontologie | | | | | 104 | 262 | 72 | 438 |
| Rocky | | | 88 | 118 | 77 | 96 | 33 | 412 |
| The Micropaleontologist | 163 | | | | | | | 163 |
| Transactions of the Kansas Academy of Science | 2107 | 611 | 307 | 263 | 236 | 293 | 48 | 3865 |
| USGS Open-File Report | 403 | 466 | 2399 | 6480 | 5060 | 726 | 243 | 15777 |
| United States Geological Survey Bulletin | 2302 | 626 | 320 | 614 | 454 | 1 | 1 | 4318 |
| United States Geological Survey Professional Paper | 596 | 721 | 733 | 465 | 227 | 71 | 54 | 2867 |
| Zoological Journal of the Linnean Society | 1165 | 121 | 363 | 483 | 487 | 638 | 392 | 3649 |
| Acta Palaeontologica Polonica | 50 | 118 | 180 | 196 | 242 | 564 | 272 | 1622 |
| Canadian Journal of Earth Sciences | | 530 | 1865 | 1981 | 1643 | 1077 | 377 | 7473 |
| Oklahoma Geology Notes | | 15 | 58 | 60 | 56 | 39 | 3 | 231 |
| Vertebrata Palasiatica | 136 | 237 | 225 | 333 | 262 | 272 | 119 | 1584 |
| Biodiversity Heritage Library | | | | | | | | 97129 |
| **Total** | | | | | | | | **277309** |

**Supplementary Table 8.** Statistics of Whole Document Set (WDS).

| | ODS | WDS | Ratio (WDS/ODS) |
|---|---|---|---|
| # Variables | 13,138,987 | 292,314,985 | 22× |
| # Evidence Variables | 980,023 | 2,066,272 | 2× |
| # Factors | 15,694,556 | 308,943,168 | 20× |
| # Distinct Features (Weight) | 945,117 | 12,393,865 | 13× |
| Documents | 11,782 | 280,280 | 23× |

**Supplementary Table 9.** Factor graph statistics in the overlapping and whole document sets. Evidence variables are those variables for which distant supervision has contributed an expectation. The scaling of evidence variables from the ODS to the WDS reflects the fact that most of the training data used by PDD derives from the PBDB data in the ODS.

| Year | Volume | Issue | Reference Title |
|---|---|---|---|
| 1993 | 13 | 3, suppl. | Ontogenetic changes in hind limb proportions within the Ghost Ranch population of Coelophysis bauri |
| 2003 | 23 | 3 | New dromomerycids (Mammalia: Artiodactyla) from the middle Miocene Sharktooth Hill Bonebed, California, and the systematics of the cranioceratinins |
| 2002 | 22 | 3 | Paleontology and stratigraphy of the Tecolotlan Basin, Jalisco, Mexico |
| 2004 | 24 | 3 | A new Miocene sperm whale (Cetacea, Physeteridae) from Virginia |
| 2003 | 23 | 3, suppl. | A preliminary Prosauropoda phylogeny with comments on Brazilian basal Sauropodomorpha |
| 2005 | 25 | 3 | A revised faunal list for the Carmel Church Quarry, Caroline County, Virginia |
| 1994 | 14 | 3, suppl. | Preliminary report on the microvertebrate fauna from the Late Cretaceous Bauru strata near Peirpolis, Minas Gerais, Brazil |
| 1993 | 13 | 3, suppl. | Sedminentology and taphonomy of the Little Houston Quarry, Morrison Formation (Upper Jurassic), northeast Wyoming |
| 1986 | 6 | 3 | |
| 2002 | 22 | 3 | A flora and faunal list of specimens recovered from the Big Pig Dig, Badlands National Park, South Dakota |

**Supplementary Table 10.** A Random Sample of PBDB References in *Journal of Vertebrate Paleontology* that Do Not Appear in the Overlapping Corpus.

| Year | Volume | Issue | Reference Title |
|---|---|---|---|
| 1905 | 22 | NULL | |
| 2006 | 312 | NULL | Comment on "The brain of LB1, Homo floresiensis" |
| 1984 | 224 | NULL | |
| 1885 | 5 | 116 | Lesquereux's Cretaceous and Tertiary Flora |
| 1991 | 251 | NULL | New fossil evidence on the sister-group of mammals and early Mesozoic faunal distributions |
| 1990 | 249 | NULL | |
| 1900 | 11 | 282 | The vertebral formula in Diplodocus Marsh |
| 1997 | 278 | NULL | A tribosphenic mammal from the Mesozoic of Australia |
| 1905 | 22 | 568 | The occurrence of ichthyosaur-like remains in the Upper Cretaceous of Wyoimng |
| 1934 | 79 | 2039 | A change of names |

**Supplementary Table 11.** A Random Sample of PBDB References in *Science* that Do Not Appear in the Overlapping Corpus.

|  | | ODS | WDS | Ratio (WDS/ODS) |
|---|---|---|---|---|
| Mention-level Candidates | Taxon | 6,049,257 | 133,236,518 | 22× |
|  | Formation | 523,143 | 23,250,673 | 44× |
|  | Interval | 1,009,208 | 16,222,767 | 16× |
|  | Location | 1,096,079 | 76,688,898 | 76× |
|  | Opinions | 1,868,195 | 27,741,202 | 15× |
|  | Taxon-Formation | 545,628 | 4,332,132 | 8× |
|  | Formation-Temporal | 208,821 | 3,049,749 | 14× |
|  | Formation-Location | 239,014 | 5,577,546 | 23× |
| Entity-level Result | Authorities | 163,595 | 1,710,652 | 10× |
|  | Opinions | 192,365 | 6,605,921 | 34× |
|  | Collections | 23,368 | 125,118 | 5× |
|  | Occurrences | 93,445 | 539,382 | 6× |
|  | Documents | 11,782 | 280,280 | 23× |

**Supplementary Table 12.** Extraction statistics for the overlapping and whole document sets. Authorities refers to distinct taxa (identified by name and, optionally, ranks and authors).

| Relation | # Annotations | Precision | Recall |
|---|---|---|---|
| Taxonomy | 933 | 97% | 39% |
| Temporal | 478 | 96% | 69% |
| Location | 655 | 92% | 36% |
| Formation | 2,271 | 94% | 21% |

**Supplementary Table 13.** Statistics of Annotations Collected and Quality Score for Each Relation

# 1 Extensions

## 1.1 Body Size Extraction

In order to extract body size estimates from biological illustrations, we need to extract the relation:

$$(Taxon, FigureName, FigureLabel, Magnification, ImageArea)$$

where $ImageArea$ is a region on the PDF with known DPI so that the actual size of the image on a printed document is known. The following table is an example of the target extracted relation.

| | | | | |
|---|---|---|---|---|
| *Vediproductus wedberensis* | Fig. 381 | 2a | X1 | 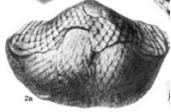 |
| *Compressoproductus compressus* | Fig. 382 | 1a | X0.8 | 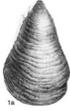 |
| *Devonoproductus walcotti* | Fig. 383 | 1b | X2.0 | 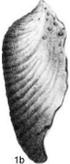 |

There were two steps in the process: (1) Image processing, and (2) text extraction. In PDD, these two components are done jointly in the same factor graph.

**Image Processing.** The goal of the image processing component is to associate each image area with a figure label. To achieve this, PDD needs to (1) detect image areas and figure labels from PDF documents, and (2) associate image areas with figure labels. Supplementary Figure 9 illustrates these two steps.

**Detection of Image Areas and Figure Labels.** The following steps were taken: (1) Edge detection; (2) Watershed Segmentation; (3) Image Dilation; and (4) Connected-component Detection (Supplementary Figure 9). Standard online-tutorials were followed, with one variant for Image Dilation. In this step, one needs to specify a parameter for dilation. Instead of specifying one value for the parameter, we tried a range of parameters and generate different versions of segmentations. PDD then trained a logistic regression classifier to choose between these segments trained on a human-labeled corpus.

**Association of Image Areas with Figure Labels.** After recognizing a set of image regions and their corresponding OCR results, PDD attempted to predict the association of figure labels and image areas, as shown in Supplementary Figure 9. Similar to relation extraction, PDD introduces a Boolean random variable for each label and image area pair. It then builds a logistic regression model using features such as the distance between label and image areas, and whether a label is nearest to an image area and vice versa.

**Text Extraction.** PDD also extracts information from text, as shown in Supplementary Figure 10. This extraction phase is similar to what was used when extracting fossil occurrence-related relations. In the name entity recognition component, PDD extracts different types of mentions, including Figure name (e.g., "Fig. 3"), Figure labels (e.g., "3a-c"), Taxon (e.g., "*B. rara*"), and magnitude (e.g., "X1"). Supplementary Figure 10 shows an example of these mentions (raw text with OCR errors). PDD then extracts relations between these mentions using the same set of features as other diversity-related relations.

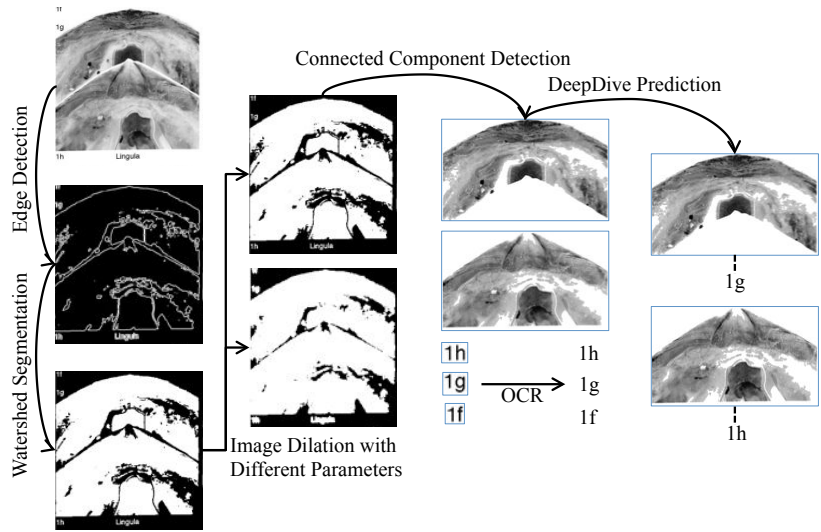

**Supplementary Figure 9.** Image Processing Component for Body Size Extraction. Note that this examples contains the illustration of a *partial* body.

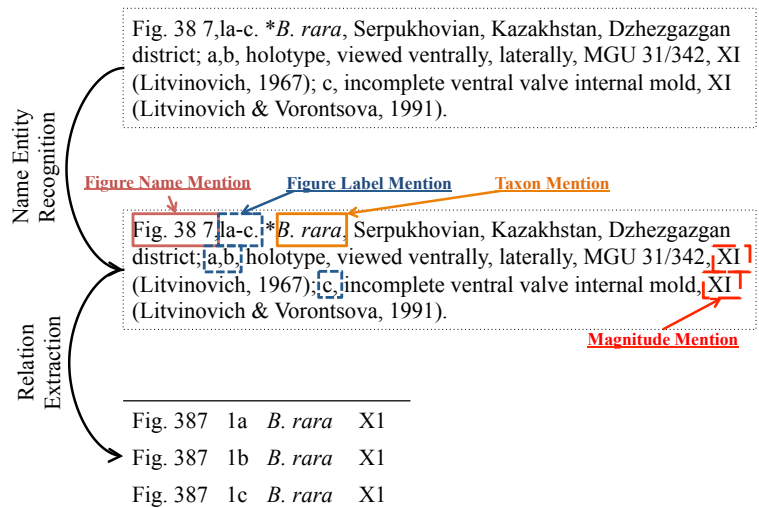

**Supplementary Figure 10.** Relation Extraction Component for Body Size Extraction.

**Joint Inference.** Both the image processing component and the text extraction component results in a factor graph populating two relations with schema

$$(FigureLabel, ImageArea)$$

and

$$(Taxon, FigureName, FigureLabel, Magnitude).$$

PDD joins these two intermediate relations to form a large factor graph to populate the target relation. Joint inference on the whole factor graph is then executed.

## 1.2 Body Size Extraction Validation

**Corpus.** Other researchers [1] recently compiled body size measurements by manually measuring illustrations and reading captions in the *Treatise on Invertebrate Paleontology*. Of the 55 volumes now accessible, humans have made measurements from part H, I, K, L, N, O, P, Q, R, S, T, U. We created from these documents the following three sets:

1. **Testing Corpus (With Ground Truth).** Part H.

2. **Testing Corpus (Without Ground Truth).** Part A, B, C, D, E, F, G, W, V.

3. **Training Corpus.** Part I, K, L, N, O, P, Q, R, S, T, U.

We used the Training Corpus to generate training data for distant supervision. We compared our results with those of human annotators using the Testing Corpus (With Ground Truth). The Testing Corpus (Without Ground Truth) shows that PDD helps to extend the body size database with new extractions that are not provided by human annotators.

**Results on Testing Corpus (With Ground Truth).** PDD is able to to achieve high precision and slightly higher recall than human when extracting body size measurements and their relations.

**Precision.** We measured the precision of PDD by randomly sampling 100 extracted instances of the target relation and manually annotate those extractions. We find that the accuracy is more than 92%.

**Recall.** We next counted the number of distinct (genus, figure name, figure label) tuples that are extracted by humans and PDD on the same set of documents. We find that human extracted 4,837 distinct tuples, and PDD extracted 5,783 distinct tuples, or 20% more. The primary reason for the increase is the complete extraction of meaurements for all parts of a figure (e.g., "1a-f"). Humans typically extract only one part.

Although selective data extraction is often a decision made for the sake of expediency and because not all images provide optimal orientations for the dimensions being targeted by a given investigation, extracting complete measurements and associated textual descriptions establishes the foundation for more complete morphometric analyses.

**Results on Testing Corpus (Without Ground Truth).** PDD is able to extract facts on documents that have not yet been processed by humans. PDD processed Parts A, B, C, D, E, F, G, V, W of the *Treatise on Invertebrate Paleontology*, which have not yet been processed for body size by [1]. PDD extracts 7K distinct (genus, figure name, figure label) tuples from these documents.

## 1.3 Multi-linguistic Extraction

**Corpus.** We followed a similar protocol as we used to collect the overlapping corpus for English documents. We identified the top-20 journals ranked by the number of journal articles in PBDB, and attempted to download articles from their web site. Access was limited to *Vertebrata Palasiatica* (Chinese), *Stuttgarter Beitrage zur Naturkunde* (German), and *Eclogae Geologicae Helvetiae* (German). A total of 1,583 Chinese journal articles and 4,393 German journal articles were obtained in this way. We used the same protocol to map these journal articles to articles in PBDB. Of these, there were 47 articles in Chinese and 56 German articles that overlapped with the PBDB.

|  | **English** | **Chinese** | **German** | **Dictionary Source** |
|---|---|---|---|---|
| Rock Formation | Formation | 组 | Formation | Manual |
|  | Clay | 石 | Ton |  |
| Temporal Interval | Late Cretaceous | 晚白垩世 | Oberkreide | Manual |
|  | Cretaceous | 白垩世 | Kreide |  |
| Location | United States | 美国 | Vereinigte Staaten | geonames.org |
| Taxon | *Aeschnidium densum* | *Aeschnidium densum* | *Aeschnidium densum* | All in Latin |

**Protocol.** We compared the extractions of PDD in the overlapping set with the PBDB extractions on the same set of documents. Our way of assessing quality is recall for the tuple

$$(Taxon, TimeInterval)$$

This tuple is language-independent because (1) taxon has unified Latin-representation in all English, Chinese, and German articles; and (2) time Intervals and their hierarchical relationships are known by PDD for all languages. To extract this tuple, PDD requires the information in all other tuples, including $(Taxon, Formation)$, $(Formation, TimeInterval)$, and $(Formation, Location)$. We selected taxa common to both PDD and PBDB, and label PDD's extraction as correct if the taxon temporal ranges overlap.

**Recall.** From the overlapping corpus, PBDB extracts $(Taxon, TimeInterval)$ tuples for 85 distinct genera in Chinese and 242 distinct genera in German. We find that PDD correctly extracts $(Taxon, TimeInterval)$ for 24 genera (28%) in Chinese and 82 (33%) genera in German. The difference between Chinese and German is caused primarily by OCR quality, even though we used commercial OCR tools for both. Chinese has lower OCR quality because of the large vocabulary in East-Asian languages.

**Precision.** Out of all 24 distinct genera in Chinese and 82 distinct genera in German articles, we find that all of them overlap with PBDB extractions in terms of their temporal interval, indicating high precision.

## 2 Specific Technical Validation

Here we describe DeepDive, the underlying system that powers PDD [2–7].

### 2.1 Probabilistic Framework

#### 2.1.1 Related Work

*Knowledge Base Construction* (KBC) has been an area of intense study over the last decade [8–19]. Within this space, there are a number of approaches.

**Rule-based Systems.** The earliest KBC systems used pattern matching to extract relationships from text. The most well known example is the "Hearst Pattern" proposed by Hearst [20] in 1992. In her seminal work, Hearst observed that a large amount of hyponyms can be discovered by simple patterns, e.g., "X, such as Y". Hearst's technique forms the basis of many further techniques that attempt to extract high quality patterns from text. In industry, rule-based (pattern-matching-based) KBC systems, such as IBM's SystemT [8, 21], have been built to develop high quality patterns. These systems provide the user a (usually declarative) interface to specify a set of rules and patterns to derive relationships. These systems have achieved state-of-the-art quality after carefully engineering effort as shown by Li et al. [21].

**Statistical Approaches.** One limitation of rule-based systems is that the developer needs to ensure that all rules provided to the system are high precision rules. For the last decade, probabilistic (or machine learning) approaches have been proposed to allow the system select between a range of a priori features automatically. In these approaches, the extracted tuple is associated with a marginal probability that it is true (i.e., that it appears in the KB). DEEPDIVE, Google's knowledge graph, and IBM's Watson are built on this approach. Within this space there are three styles of systems:

- **Classification-based Frameworks** Here, traditional classifiers assign each tuple a probability score, e.g., naïve Bayes classifier, and logistic regression classifier. For example, KnowItAll [12] and TextRunner [13, 14] uses naïve Bayes classifier, and CMUs NELL [16, 17] uses logistic regression. Large-scale systems typically use these types of approaches in sophisticated combinations, e.g., NELL or Watson.

- **Maximum a Posteriori (MAP)** Here, the probabilistic approach is used but the MAP or Most likely world (which do differ slightly) is selected. Notable examples include the YAGO system [15],which uses a PageRank-based approach to assign a confidence score. Other examples include the SOFIE [10] and Prospera [11], which use an approach based on constraint satisfaction.

- **Graphical Model Approaches** The classification-based methods ignore the interaction among predictions, and there is a hypothesis that modeling these correlations yields higher quality systems more quickly. A generic graphical model has been used to model the probabilistic distribution among all possible extractions. For example, Poon et al. [19] used Markov logic networks (MLN) [22] for information extraction. Microsoft's StatisticalSnowBall/EntityCube [18] also uses an MLN-based approach. A key challenge with these systems is scalability. For example, Poon et al. was limited to 1.5K citations. Our relational database driven algorithms for MLN-based systems are dramatically more scalable [3].

### 2.1.2 Calibrated Probabilities

DEEPDIVE takes a Bayesian probabilistic approach to KBC by treating OCR, NLP, image processing, and feature recognition as one joint probabilistic inference problem in which all predictions are modeled as a factor graph (Fig. S3). This probabilistic framework ensures all facts that are produced by DEEPDIVE are associated with a marginal probability.[1] These marginal probabilities are meaningful in DEEPDIVE (i.e., they should correspond to the actual probabilities of a fact beig correct), which provides a mehcanism for evaluation and an aid to improving the system.

**Calibration.** In DEEPDIVE, *calibration plots* are used as a way to summarize the overall quality of the KBC results. Ideally, the probability associated with a given fact in DEEPDIVE should equal the empirical probability that this fact is correct (i.e., an extraction with a probability 0.95 should be correct with a 95% of the time when inspected in the original source). Because DEEPDIVE uses a joint probability model, any set of predictions can be assigned a marginal probability. Queries can then be against the model to help determine where a model needs improvement.

Supplementary Figure 11 and Supplementary Figure 12 show calibration plots for the ODS and the WDS presented in the main text. We will use Supplementary Figure 11(1) as an example, which is the target relation Taxonomy in the ODS. A calibration plot contains three components: (a) Accuracy, which measures the test-set accuracy of a prediction with a certain probability; (b) # Predictions (Testing Set), which measures the number of extractions in the test set with a certain probability; and (c) # Predictions (Whole Set), which measures the number of extractions in the whole set with certain probability. The difference between test set and whole set is that the former has training labels for each random variable. Results are summarized as histograms, and empirically we find that a bin of size of 0.1 is usually sufficient to understand the behavior of the system.

---

[1] Cox's theorem asserts (roughly) that if one uses numbers as degrees of belief, then one must either use probabilistic reasoning or risk contradictions in a reasoning system, i.e., probabilistic reasoning is the only sound system for reasoning in this manner [23].

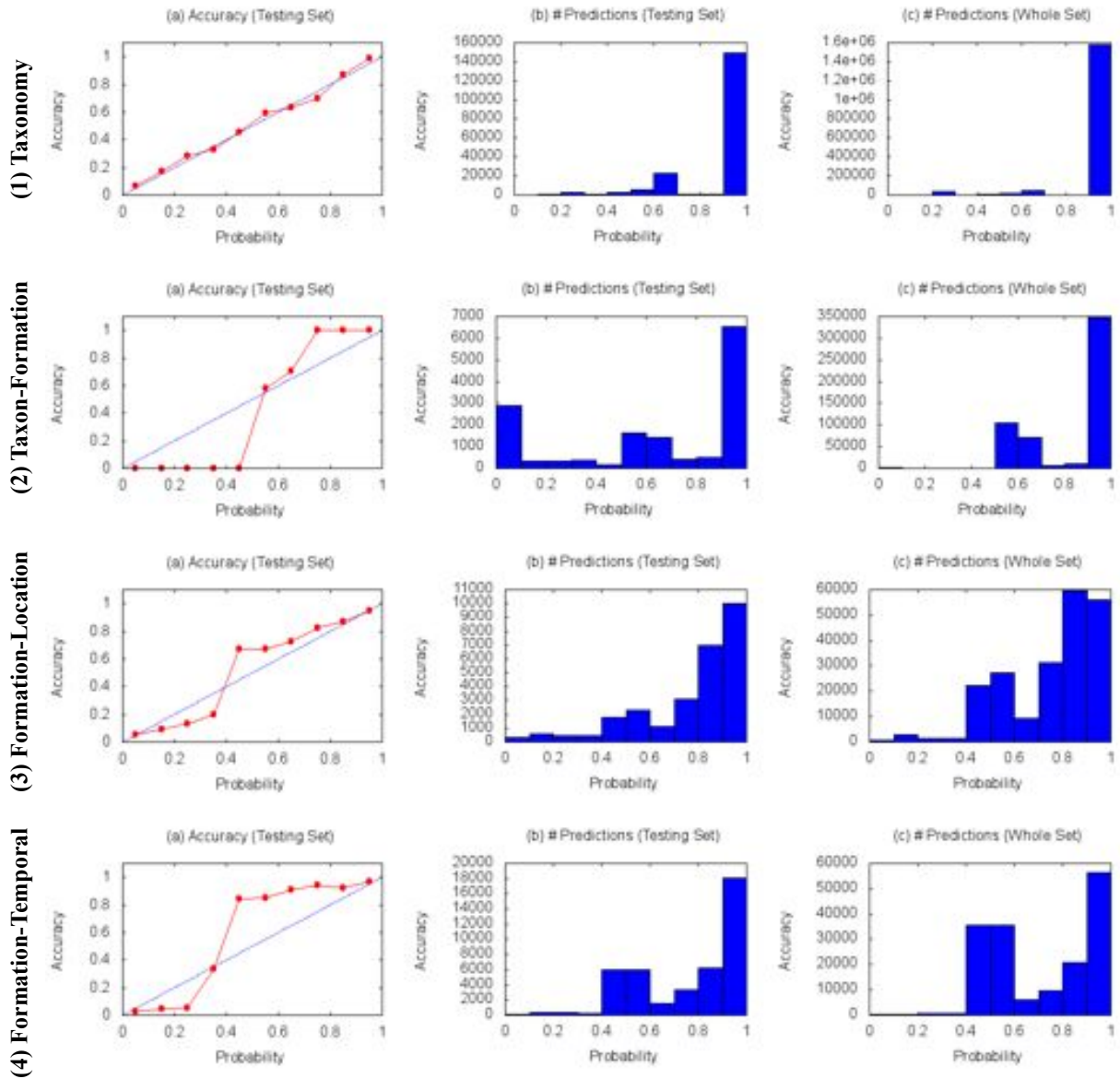

**Supplementary Figure 11.** Calibration Plots for All Relations on Overlapping Corpus

### Using Calibration Plots

**(a) Accuracy.** If the accuracy curve is similar to the ideal (0,0)-(1,1) line, it means that a probability produced by the system matches the *test-set accuracy*. For example, Supplementary Figure 11(1) shows a reasonably good curve for calibration. Differnces in these two lines can be caused by (1) inefficient training data or a small testing corpus, and/or (2) bad mixing behavior of the sampler or other software bugs. For example, Supplementary Figure 12(2,3,4) shows a much better calibration behavior than Supplementary Figure 11(2,3,4), primarily because the former is based on the whole corpus, which has more training data and a larger testing set.

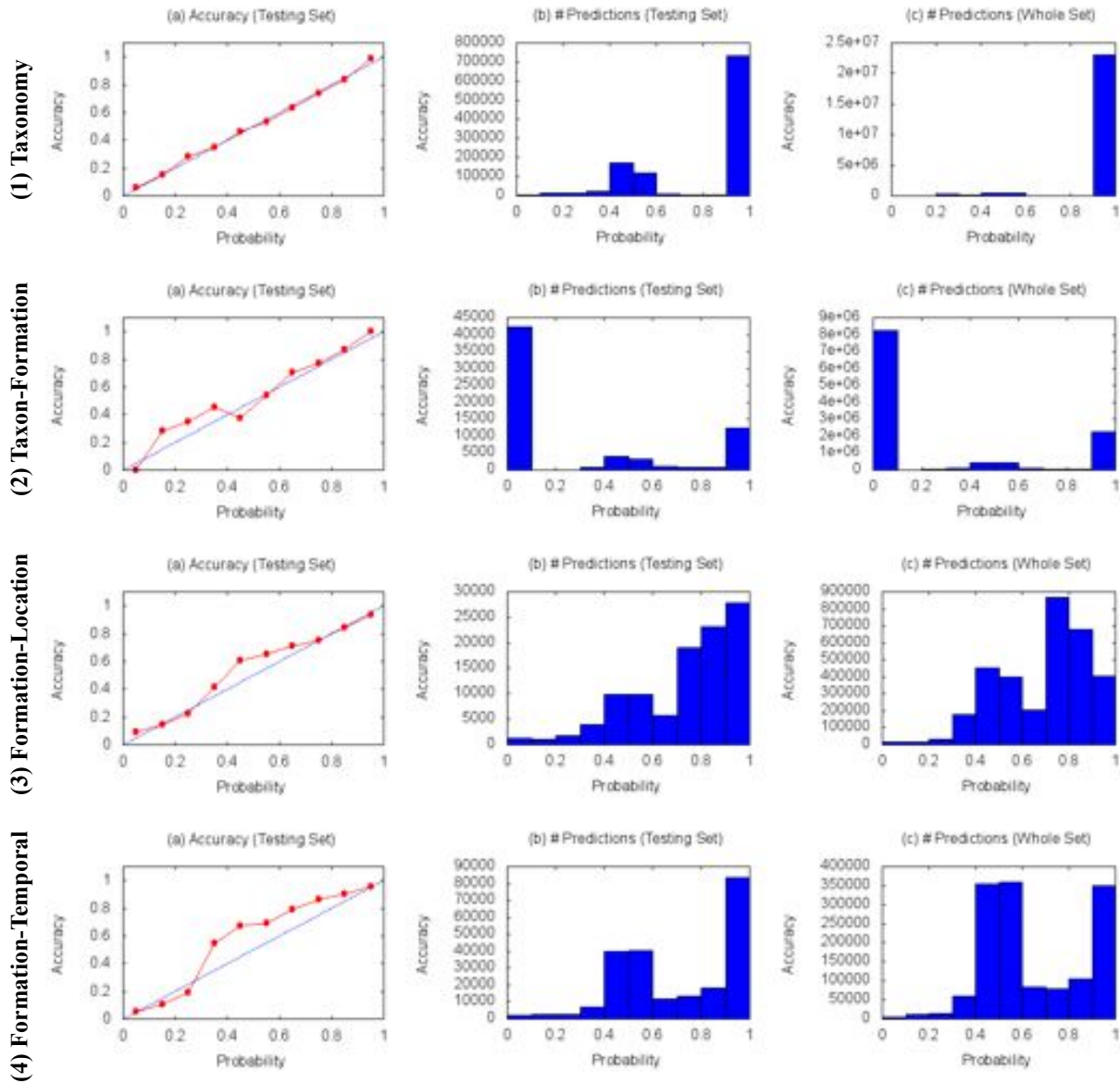

**Supplementary Figure 12.** Calibration Plots for All Relations on Whole Corpus

**(b) # Predictions (Testing Set).** Ideally, the # Predictions histogram should have a "U" shape. That is, most of the data are concentrate at high probability (where we are confident it is correct) and low probability (where we are confident it is incorrect). Large numbers of predictions with a probability approximately 0.5 means that the system has little information about how to classify these extractions. This implies that more features could be defined to resolve uncertainty. For example, Supplementary Figure 11(2) shows a U-shape curve with some masses around 0.5-0.6. The shape of the histogram relies on the ratio between the number of positive examples and negative examples. When the number of positive examples dominates negative examples and there is a bias term, it is possible that there are very small amount extractions with a probability near 0. Supplementary Figure 11(1,3,4) illustrate this phenomenon.

**(c) # Predictions (Whole Set).** This histogram is similar to (b), but illustrates the behavior of scaling the system to a set of documents for which we do not have any training examples. Usually we hope that (c) has a similar shape to (b).

**Usage.** The above techniques have proven critical to debugging and improving the quality of PDD. In response to low confidence, a user can provide labeled examples, which allows the system to learn weights that yield higher confidence. Additionally, a user may write logical inference rules that provide ways of improving quality, which is a key component of all statistical relational approaches.

## 2.2 Declarative Interface for Joint Inference and Rich Features

### 2.2.1 Related Work

Here we survey recent efforts that focus on how to improve the quality of a KBC system.

**Rich Features.** Different researchers have recently noted the importance of combining and using a rich set of features and signals to improve the quality of a KBC system. Two famous efforts, the Netflix challenge [24], and IBM's Watson [25], which won the Jeopardy gameshow, have identified the importance of features and signals:

> *Ferrucci et al. [25]: For the Jeopardy Challenge, we use more than 100 different techniques for analyzing natural language, identifying sources, finding and generating hypotheses, finding and scoring evidence, and merging and ranking hypotheses. What is far more important than any particular technique we use is how we combine them in DeepQA such that overlapping approaches can bring their strengths to bear and contribute to improvements in accuracy, confidence, or speed.*

> *Buskirk [24]: The top two teams beat the challenge by combining teams and their algorithms into more complex algorithms incorporating everybody's work. The more people joined, the more the resulting team's score would increase.*

In both efforts, the rich set of features and signals contributed to the high-quality of the corresponding system. Other researches have found similar phenomena. For example, Mintz et al. [26] finds that although both surface features and deep NLP features have similar quality for relation extraction tasks, combining them achieves a significant improvement over using either one in isolation. Similar "feature-based" approaches are also used in other domains (e.g., Finkel et al. [27] uses a diverse set of features to build a NLP parser with state-of-the-art quality). In our own work [28], we have also found that integrating a diverse set of deep NLP features can improve a table extraction system significantly.

**Joint Inference.** Another recent trend in building KBC system is to take advantage of *joint inference* [5, 19, 28–33]. Different from traditional models [34], such as logistic regression or SVM, joint inference approaches emphasize learning multiple targets simultaneously. For example, Poon et al. [19, 31] find that learning segmentation and extraction in the same Markov logic network significantly improves the quality of information extraction. Similar observations have been made by Min et al. [29] and McCallum [30]. Our recent work also show the empirical improvement of joint inference on the diverse set of tasks, including relation extraction [5] and table extraction [28].

**Deep Learning and Joint Inference.** A recent emerging effort in the machine learning community is to build a fully-joint model for NLP tasks [32, 33]. The goal is to build a single joint model from the lowest level (e.g., POS tagging) to the highest level (e.g., semantic role labeling). The PDD system is built in a similar spirit that attempts to build a joint model for low-level tasks (e.g., OCR), to high-level tasks (e.g., cross-document inference of relation extraction).

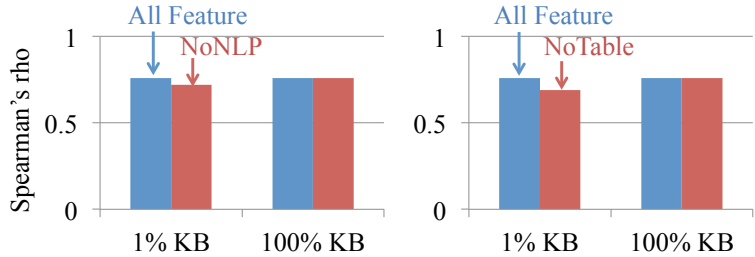

**Supplementary Figure 13.** Lesion Study of Deep NLP Features and Table Recognition

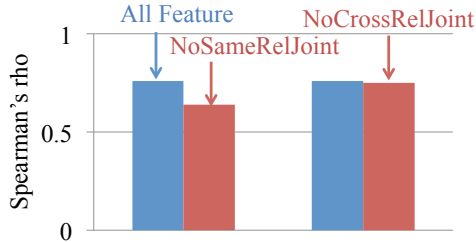

**Supplementary Figure 14.** Lesion Study of Joint Inference

### 2.2.2 The DeepDive Approach and the Impact of Rich Features and Joint Rules

DEEPDIVE uses joint inference rules and rich features. In this section, we test that these features and rules are important to PDD's quality by conducting a lesion study.

**Protocol.** All experiments were run on the overlapping corpus as described in the main text. We produced variants of PDD by removing features/rules and all components that rely on the output of the removed feature/rule. We summarize the quality of PDD by computing Spearman's rho for first differences in genus-level biodiversity (as in Fig. 1).

**Features.** The PDD feature extraction phase extracts a set of features, including deep linguistic features, e.g., dependency parsing results, and vision-based features (e.g., a simple table extractor based on Hough Transform). To study their impact, we conduct lesion study by sequentiallydisabling these features.

**Deep NLP Features.** Supplementary Figure 13(a) shows the impact of removing NLP features (e.g., dependency path). If we use the whole PBDB is used, dropping these Deep NLP features does not have a significant effect on Spearman's rho. However, if the knowledge base used for training is reduced to 1% of it s size, then dropping NLP features results in a decrease of Spearman's rho from 0.72 from 0.82.

**Vision-based Table Recognition.** PDD contains a table recognition component to detect tables using vision-based features (e.g., Hough Transform). When disabling this component and using the 1% PBDB for distant supervision, PDD achieves a Spearman's rho of 0.69. This drop is the effect of decreased recall of data in tables.

**Joint Inference Rules.** PDD contains a set of factors for joint inference among random variables, as shown in Fig S3. We study their impact on two types of joint inference rules: (1) joint inference within one relation; and (2) joint inference across different relations (Supplementary Figure 14).

**Joint Inference for Same Relations.** Disabling all joint inference rules results in a Spearman's rho of 0.64, even when using the whole PBDB knowledge base. This is a marked decline from the Spearman's rho of 0.82 obtained when these rules are enabled. This large decline in quality is caused by the fact that jointly infering the values of random variable results in much higher-quality predictions. For example, assume that we have three candidate facts that Tsingyuan Formation has the age (1) Carboniferous ,(2) Namurian, and (3) Kungurian. In the current PDD system, the higher confidence for Carboniferous will also boost its confidence for Namurian (because of containment), and decrease its confidence for Kungurian (because Kungurian is so much younger than Carboniferous). This type of joint inference between random variables help PDD to produce result with higher recall (by boosting confidence to cross the imposed 0.95 threshold) and precision (by eliminating wrong predictions).

**Joint Inference across Relations.** The current PDD system has three joint inference rules across different relations (e.g., one geologic formation entity mention cannot be concurrently a location mention). We disable these rules and show in Supplementary Figure 14 that it does not have a large impact to the overall quality. This implies that the current PDD system is quite modular across different relations. This means that different types of relations can be decoupled and applied to other related applications (e.g., for biology or geology).

## 2.3 Scalability and High Performance Statistical Inference and Learning

### 2.3.1 Related Work

There is an emerging trend in both industry and academia to support statistical inference and learning, and we survey these efforts in this section.

**Hardware Efficiency.** One line of research tries to speed-up statistical inference and learning by better taking advantage of modern hardware and clusters. For example, many industrial database vendors have integrated statistical analytics components into their product. For example, Oracle's ORE [35], Pivotal's MADlib [36], and IBM's SystemML [37]. These systems provide functionalities like logistic regression and collapsed Gibbs sampling for topic modeling on their data management systems. There are also efforts to design new data processing framework instead of relying on the traditional database systems. Indeed, most data processing frameworks developed in the last few years are designed to support statistical analytics including Mahout [38] for Hadoop, MLI for Spark [39], GraphLab [40], GraphChi [41], and Delite [42, 43]. These systems have been shown to increase the performance of corresponding statistical analytics tasks significantly.

**Statistical Efficiency.** One key difference between statistical inference and learning with traditional SQL-like analytics is that different ways of executing the same tasks usually lead to different speed when converging to the same quality. Therefore, another line of related work, mainly contributed by the mathematical optimization and machine learning community, is to design more efficient algorithms for statistical inference tasks. One of the recent trends is to design lock-free algorithms that can be executed on the emerging multi-socket multi-core machines with high parallelism [3, 44–47]. For example, Tsitsiklis et al. [44] proves asymptotic convergence for a parallel coordinate descent algorithm, and Bradley et al. [47] proves the convergence rate and theoretical speedups for parallel stochastic coordinate descent. Our own work [3, 46] proves the convergence of lock-free execution for stochastic gradient descent and stochastic coordinate descent.

### 2.3.2 The DeepDive Approach and The Performance of PDD

**The DeepDive Approach.** The statistical inference and learning engine in DEEPDIVE [4] is built upon the challenge of designing a high-performance statistical inference and learning engine on a single machine [4, 6, 7, 46]. Compared to traditional work, the main novelty of DEEPDIVE is that it considers *both* hardware efficiency and statistical efficiency for executing an inference and learning task.

**Hardware Efficiency.** DeepDive takes into consideration the architecture of modern non-uniform memory access (NUMA) machines. A NUMA machine usually contains multiple nodes (sockets), where each sockets contains multiple CPU cores. To achieve high hardware efficiency, it is useful to decrease the communication across different NUMA nodes.

**Statistical Efficiency** Pushing hardware efficiency to the extreme might cause statistical efficiency to suffer because the lack of communication between nodes could decrease the rate of convergence of a statistical inference and learning algorithm. DeepDive takes advantage of theoretical results of model averaging [45] and lock-free execution [7, 46].

**Performance of Statistical Inference and Learning.** DeepDive enables PDD's ability to run statistical inference and learning efficiently. For example, on the whole corpus, the factor graph contains more than 0.2 billion random variables and 0.3 billion factors. On this factor graph, DeepDive is able to run Gibbs sampling on a machine with 4 sockets (10 core per sockets), and we find that we can generate 1,000 samples for all 0.2 billion random variables in 28 minutes.